\documentclass[aps,prb,twocolumn,reprint,numerical,groupedaddress,floatfix,citeautoscript]{revtex4-1}
\usepackage{amsmath,amssymb}
\usepackage{graphicx}
\usepackage{dcolumn}
\usepackage{bm}
\usepackage{anysize}
\usepackage[outercaption]{sidecap}
\usepackage[pass,letterpaper]{geometry}
\usepackage{color}
\usepackage[caption=false]{subfig}
\newcommand{\subfigimg}[3][,]{
	\setbox1=\hbox{\includegraphics[#1]{#3}}
	\leavevmode\rlap{\usebox1}
	\rlap{\hspace*{-0.35cm}\raisebox{\dimexpr\ht1-0.6\baselineskip}{#2}}
	\phantom{\usebox1}
}

\begin{document}

\title{Non-diffusive Lattice Thermal Transport in Si-Ge Alloy Nanowires}
\author{M. Upadhyaya and Z. Aksamija}
\affiliation{Department of Electrical and Computer Engineering, University of Massachusetts-Amherst, Amherst, MA 01003, USA}
\email{zlatana@engin.umass.edu}

\begin{abstract}

We present a calculation of the lattice thermal conductivity of Si-Ge nanowires (NWs), based on solving the Boltzmann transport equation by the Monte Carlo method of sampling the phonon mean free paths. We augment the previous work with the full phonon dispersion and a partially diffuse momentum-dependent specularity model for boundary roughness scattering. 
We find that phonon flights are comprised of a mix of long free-flights over several $\mu$m interrupted by bursts of short flights, resulting in a heavy tailed distribution of flight lengths, typically encountered in L\'{e}vy walk dynamics. Consequently, phonon transport in Si-Ge NWs is neither entirely ballistic nor diffusive; instead, it falls into an intermediate regime called superdiffusion where thermal conductivity scales with the length of the NW as $\kappa \propto L^{\alpha}$ with the exponent of length dependence $\alpha\approx 0.33$ over a broad range of wire lengths 10 nm$<L<$10 $\mu$m regardless of diameter and roughness. We conclude that thermal conductivity in Si-Ge alloy NWs is length-dependent up to 10 $\mu$m and therefore can be tuned for thermoelectric applications.
\date{\today}
\end{abstract}
\maketitle

\section{Introduction} \label{Introduction}
Si-Ge alloys and their nanostructures have attracted attention for thermoelectric (TE) application due to their ability to achieve high figure of merit (ZT) at temperatures much above room temperature, as well as the possibility of integrating such nanostructures in other Si-based technologies, such as microelectronics. Semiconductor alloys often make good thermoelectrics because they closely resemble the \emph{phonon crystal electron glass (PGEC)} concept\cite{SnyderNMAT08}: they retain an ordered crystal structure but introduce disorder through mass variation. Mass disorder dramatically reduces lattice thermal transport while affecting electrons to a much lesser degree. The success of Si-Ge alloys for TE applications stemmed primarily from the dramatic order-of-magnitude reduction of the lattice thermal conductivity in bulk alloys due to strong mass-difference scattering of phonons with the constituent components of the alloy. Subsequently, similar increase in the thermoelectric figure of merit was observed in thin silicon wires \cite{HochbaumNAT08}; the improvement was credited primarily to the dramatic reduction of lattice thermal conductivity arising from the strong scattering of phonons with roughness at the boundaries of the wires \cite{MartinPRL09}. 

Theoretical calculations predicted that, if the reduction of thermal conductivity due to alloy scattering could be combined with the effect of boundary scattering, then Si-Ge-based nanowires (SiGe NWs) could lead to even more dramatic improvements of ZT \cite{ShiAPL09}. Subsequent MD simulations showed that, while the maximum reduction of thermal conductivity in bulk Si-Ge alloys exceeds one order of magnitude, the reduction in alloy NWs is only a factor of 5 \cite{ChenAPL09}. Several measurements of thermal conductivity in thin SiGe NWs with rough boundaries were performed \cite{KimAPL10,YinAPL12,LeeNL12} and confirmed the weak diameter dependence of thermal conductivity in Si-Ge NWs, indicating that thermal transport was dominated by alloy (mass-difference) scattering, even at low Ge concentrations. Hsiao et al. studied length dependence in SiGe NWs and found a linear trend, attributed to ballistic transport, with a clear transition to diffusive regime at lengths exceeding 8.3 $\mu$m \cite{HsiaoNatNano13}. In bulk SiGe alloys, on the other hand, Vermeersch et al. \cite{VermeerschPRB15} have argued that the phonon transport, on time scales up to 2 ns, shows clear signs of superdiffusion.

In this paper, we study length-dependent thermal transport in SiGe NWs. In Sec.~\ref{section2} we present the details of our model used to study thermal transport in SiGe NWs using a full-band description of the phonon dispersion together with a momentum-dependent specularity description of scattering at the partially diffuse boundaries of the wire. We apply the Monte Carlo (MC) technique to sample phonon mean-free-paths in the Si-Ge NW in order to fully capture the interaction between the strong alloy scattering inside the wire and partially diffuse roughness scattering at its boundaries. In Sec.~\ref{section3} we discuss our results, and show that phonons exhibit a mix of rare micron-long free flights, interspersed with diffuse scattering due to alloy mass disorder and interactions with the rough boundaries. Collectively, this leads to a heavy tailed distribution of phonon mean-free-paths (MFPs), typically found in L\'{e}vy walks \cite{BlumenPRA89,BarthelemyNature08}. This unique feature fundamentally changes transport in Si-Ge NWs and causes superdiffusion, which is evidenced by a sub-linear $\kappa(L) \propto L^{1/3}$ length scaling over a broad range of wire lengths (10 nm$<L<$10 $\mu$m) and a complete absence of a direct ballistic-to-diffusive transition. Similarly, we find the time evolution of mean-square energy displacement to be superlinear ($\sigma^2(t)\propto t^\beta$, with $\beta$=1.34), confirming superdiffusive transport of phonons in Si-Ge NWs. Finally, in Sec.~\ref{sec:conclusion} we conclude with a brief summary and a few final remarks.

\section{Model}\label{section2}
Semiconductor NWs are typically grown using the VLS procedure \cite{LiAPL03_1}, producing a circular geometry which leads to the usual Casimir limit $\tau_B^{-1}\propto v_s/D$ in the case where boundary scattering is independent of angle \cite{XieAPL14}. However, no closed-form solution to the BTE can be found for the case where there are both partially specular boundary scattering and strong internal (umklapp+alloy+defect) scattering present \cite{XiePCCP13}, and Mathiessen's rule is often used to combine the rates due to boundary scattering with the intrinsic mechanisms \cite{WangAPL10}. The closed-form solution used in the planar (membrane or thin film) case \cite{AksamijaPRB13,SellanJAP10} is not valid here because the distance from each point on the interior of the circular wire to the rough surface depends on both the angle (determined by the phonon group velocity vector) and the point of origin, in contrast to the planar case. 
In addition, the specularity of the boundary depends not only on the roughness, but also on the angle of incidence between the phonon and the local boundary normal \cite{AksamijaPRB10,SofferJAP67}. 

To overcome these challenges, we turn to the phonon Monte Carlo (pMC) technique, which has been widely used to solve the phonon Boltzmann transport equation (pBTE) \cite{MazumderJHT01,ChenJHT05,LacroixAPL06,JengJHT08,Randrianalisoa08,PeraudPRB11,RamayyaPRB12}. The pMC allows us to sample the phonon lifetimes \cite{McGaugheyAPL12} and find the combined phonon lifetime in the presence of both intrinsic scattering (from anharmonic phonon-phonon and mass difference interactions inside the wire) and partially diffuse boundary scattering at the rough surface in circular wires \cite{MartinPRL09}. We capture the anisotropy of thermal transport in Si-Ge NWs \cite{LiJAP13} due to phonon focusing effects \cite{AksamijaPRB10}, by expanding the pMC algorithm to include the full phonon dispersion \cite{KukitaJAP13,MeiJAP14}. We use Weber's Adiabatic Bond Charge model \cite{WeberABC2} to efficiently compute the full phonon dispersion of bulk silicon and germanium, and then combine them in the virtual crystal approximation (VCA). This approach has been shown to accurately capture the vibrational frequencies and group velocities of phonons in the alloy \cite{LarkinJAP13}, as well as the thermal conductivity over a broad range of compositions \cite{UpadhyayaJMR15,KhatamiPRApp16}. 

In the phonon Monte Carlo algorithm, an ensemble of phonons is initialized according to the Bose-Einsten distribution \cite{McGaugheyAPL12}. Then the free-flight time until scattering of each phonon is determined by first selecting a random number $r_{int.}$, uniformly distributed between 0 and 1, and sampling the lifetime according to $t_{int.}=-ln(r_{int.}) \tau_{int.}(\vec q)$ \cite{MaurerAPL15}. The phonon lifetime $(\tau_{int.}\vec q)$ combines all the intrinsic scattering mechanisms, including anharmonic 3-phonon interactions, impurity, isotope, and alloy mass-difference scattering (the details of which can be found in the Appendix). Once the phonon free-flight time is determined, each phonon travels along the propagation direction given by its group velocity $\vec v_g(\vec q)$, until scattering at the end of the "free flight" at $t_{int.}$ or until it encounters a boundary or contact, whichever occurs first. 

In our approach, boundary roughness scattering is characterized through a momentum-dependent specularity \cite{SofferJAP67} $p(\vec q)=\exp(-\langle \phi^2\rangle)$, where $\phi(\vec q,\vec r)=2\vec q\cdot{\hat s}z(\vec r)=2q z(\vec r) \cos{\Theta_B}$ is the phase difference between the incoming wave and the outgoing specularly reflected wave at point $\vec r$. The surface normal unit vector at this point $\vec r$ is $\hat s$. We assume that the surface height $z(\vec r)$ is a random function of position on the rough boundary $\vec r$ with a Gaussian distribution, so that $\langle z\rangle=0$ and $\langle z^2\rangle=\Delta^2$, where $\Delta$ is the rms height of the surface roughness \cite{GoodnickPRB85}. When a phonon reaches the rough boundary, another random number $r_{spec.}$ again uniformly distributed between 0 and 1, is used to select between a specular (mirror-like reflection about the surface normal $\hat s$) and a diffuse (direction after leaving surface is randomized) scattering. 
If $r_{spec.}<p(\vec q)$, then the boundary interaction is specular and the phonon is simply reflected at the boundary by flipping its momentum about the boundary normal $\vec q_{final}=\vec q_{init.}-2 \vec q_{init.}\cdot \hat{s}$. Otherwise, the interaction is diffuse: the phonon path is terminated and the boundary scattering time $t_B$ is recorded as the time at which the diffuse scattering occurred and the process is restarted. We also capture the quasi-ballistic contribution arising from phonons which reach the contacts (at time $t_{C}$) before scattering internally or at the boundaries. When all the phonon flights are terminated in either internal or boundary scattering or at a contact, thermal conductivity is computed from the average 
\begin{equation}
\kappa=\frac{1}{N_{\vec q} N_i}\sum_{\vec q, i} v_g^2(\vec{q}) \min \{t_{int.},t_B,t_C\} C(T,\vec q)
\end{equation}
where $C(T,\vec{q})$ is the modal volumetric heat capacity \cite{McGaugheyAPL12} and $N_{\vec q}$, $N_i$ are the number of phonons and iteration in the simulation, respectively, both being typically 100,000.


\section{Results and Discussion}\label{section3}

\subsection{Comparision of our results to experimental data}

\begin{figure*}
	\centering
	\subfloat[]{\label{fig:kappa06}%
		\includegraphics[width=0.25\textwidth]{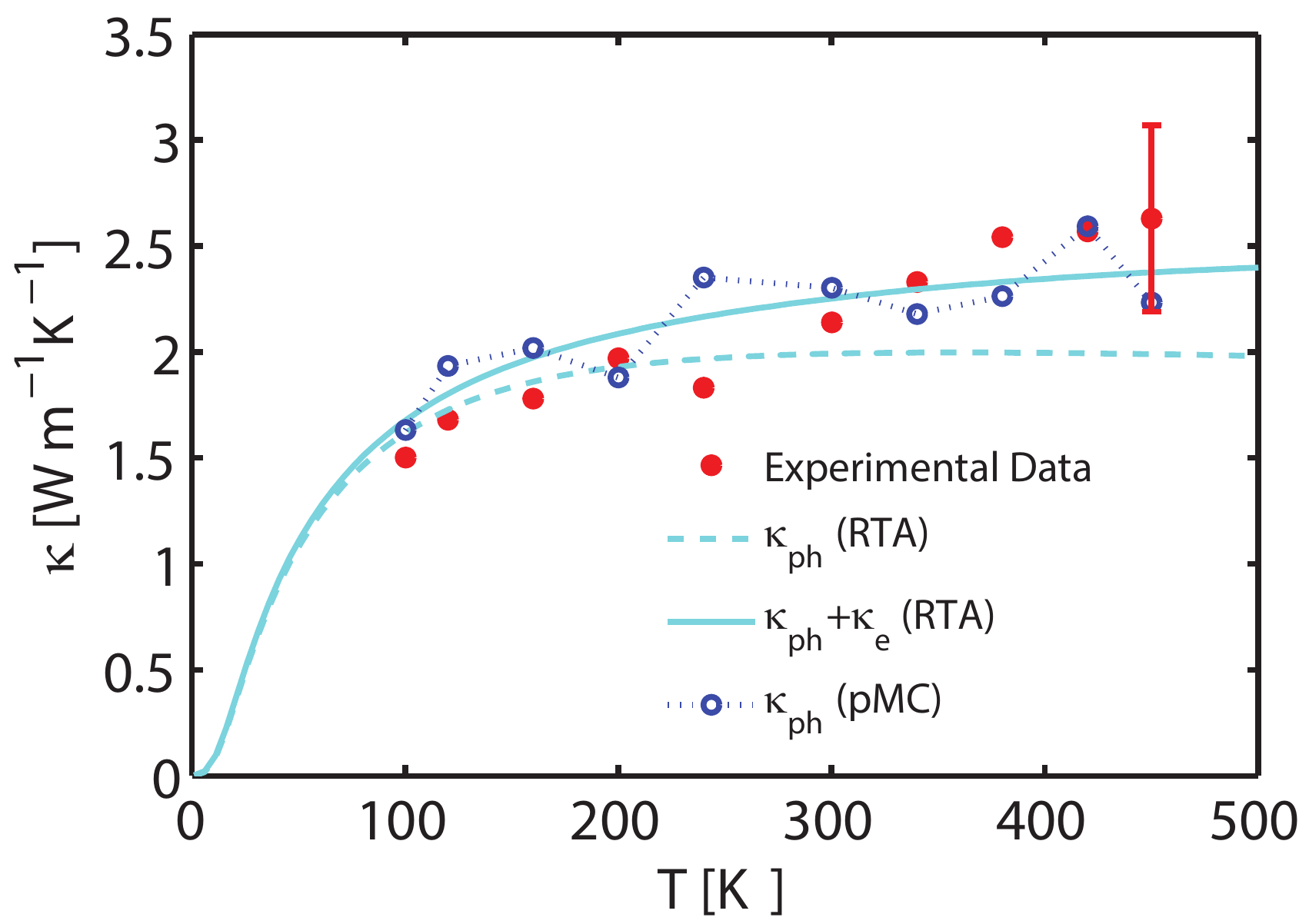}}
	\subfloat[]{\label{fig:kappa08}%
		\includegraphics[width=0.25\textwidth]{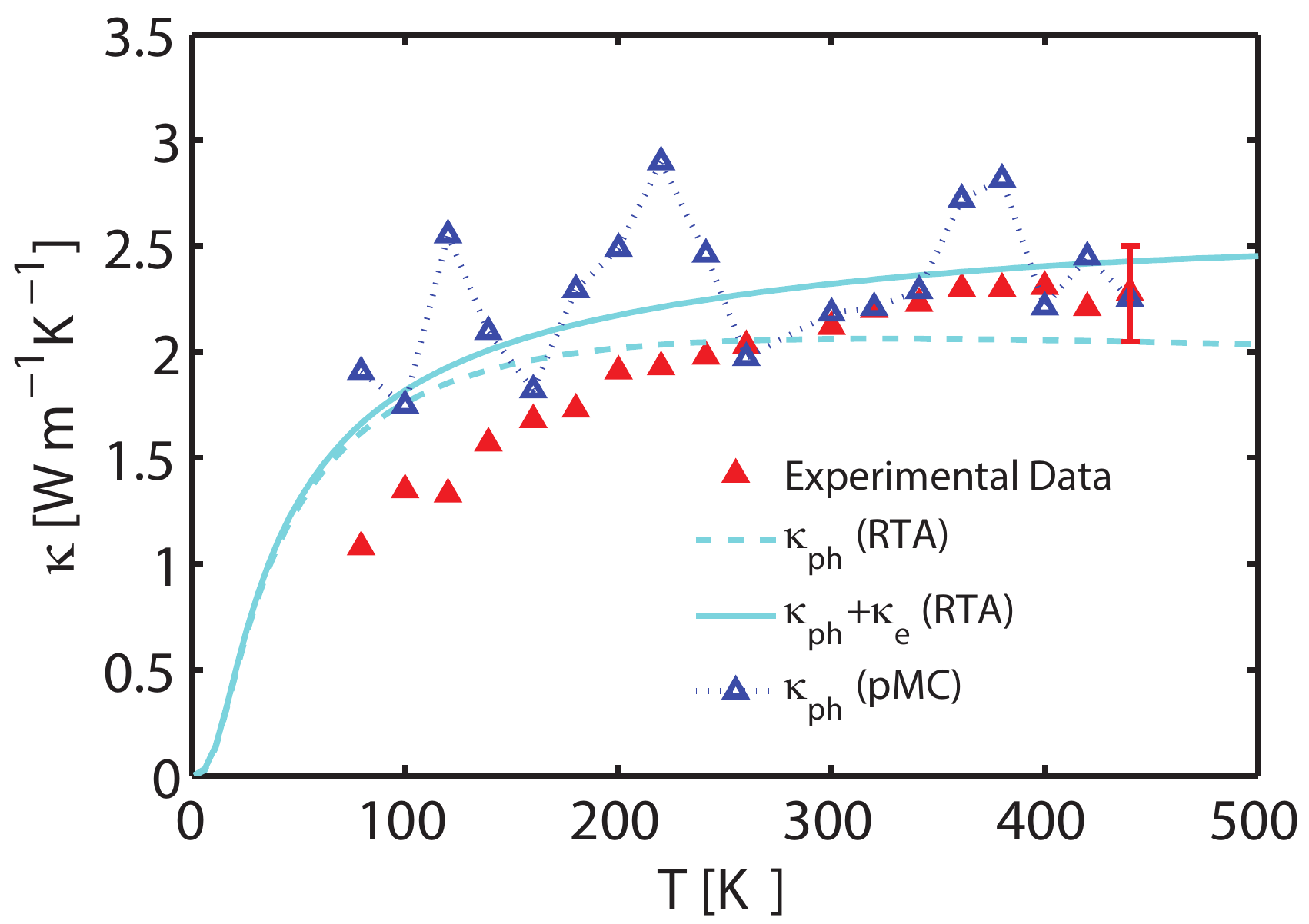}}
	\subfloat[]{\label{fig:kappa10}%
		\includegraphics[width=0.25\textwidth]{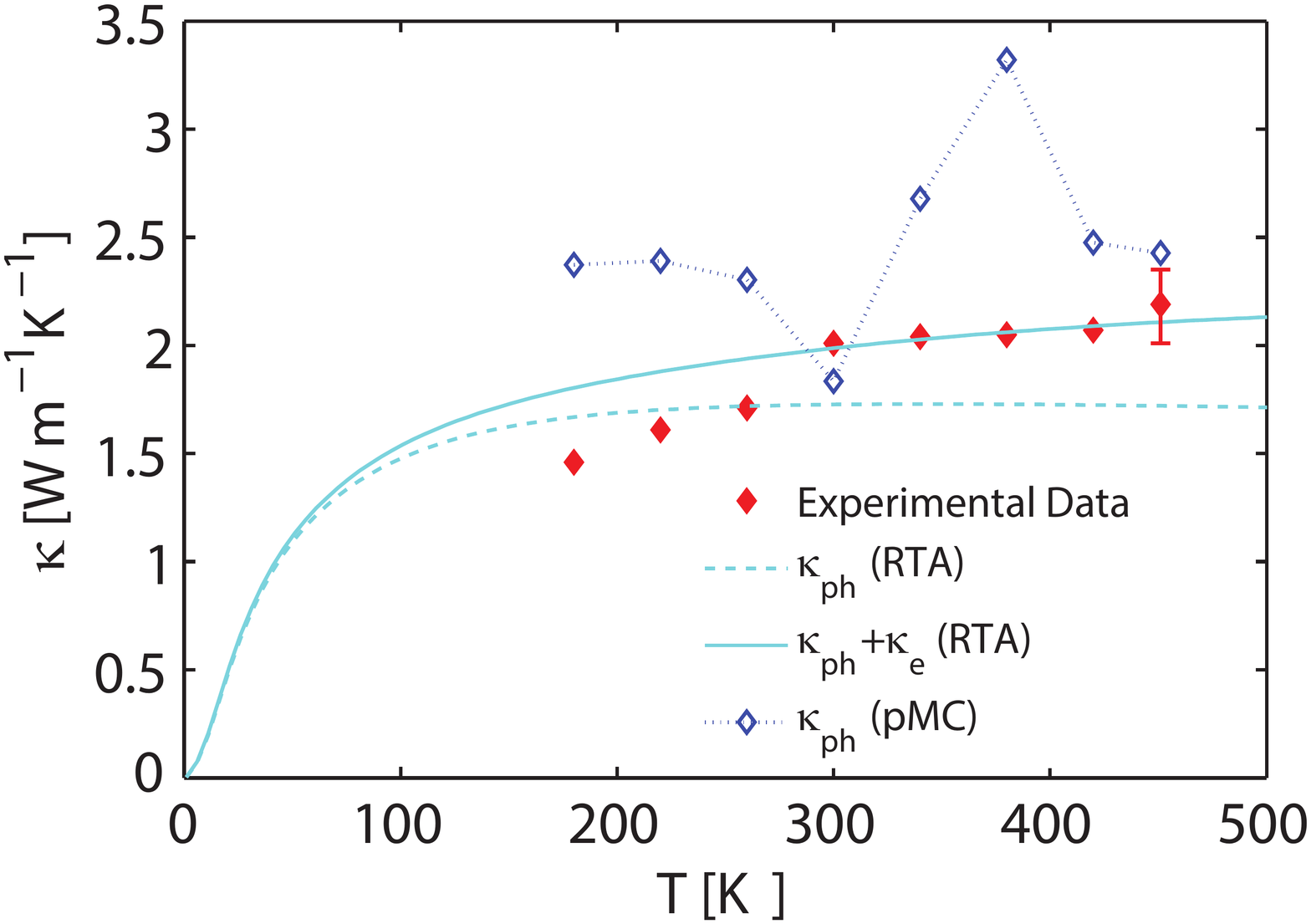}}
	\subfloat[]{\label{fig:kappa19}%
		\includegraphics[width=0.25\textwidth]{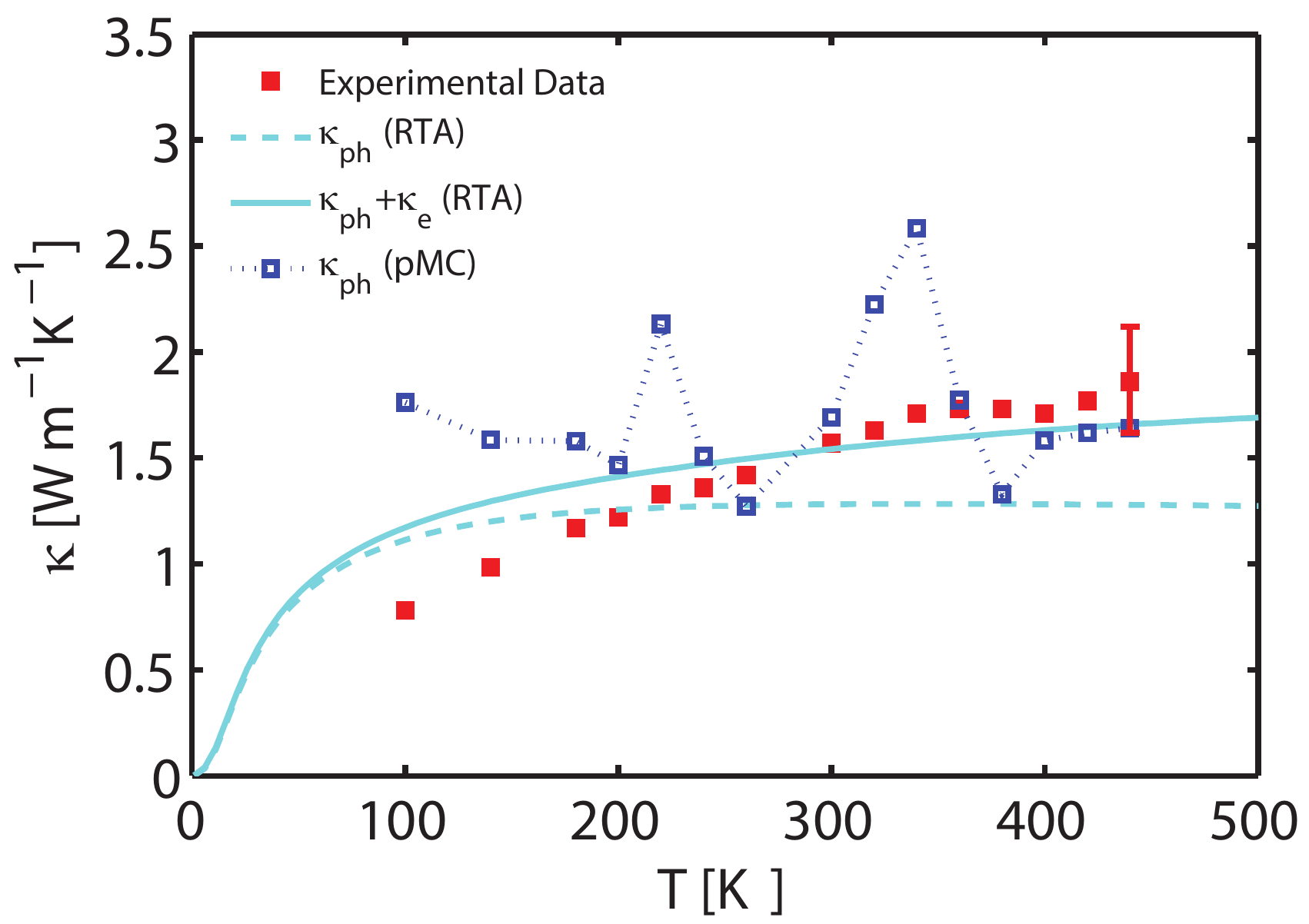}}
	\newline
	\subfloat[]{\label{fig:kappa26}%
		\includegraphics[width=0.25\textwidth]{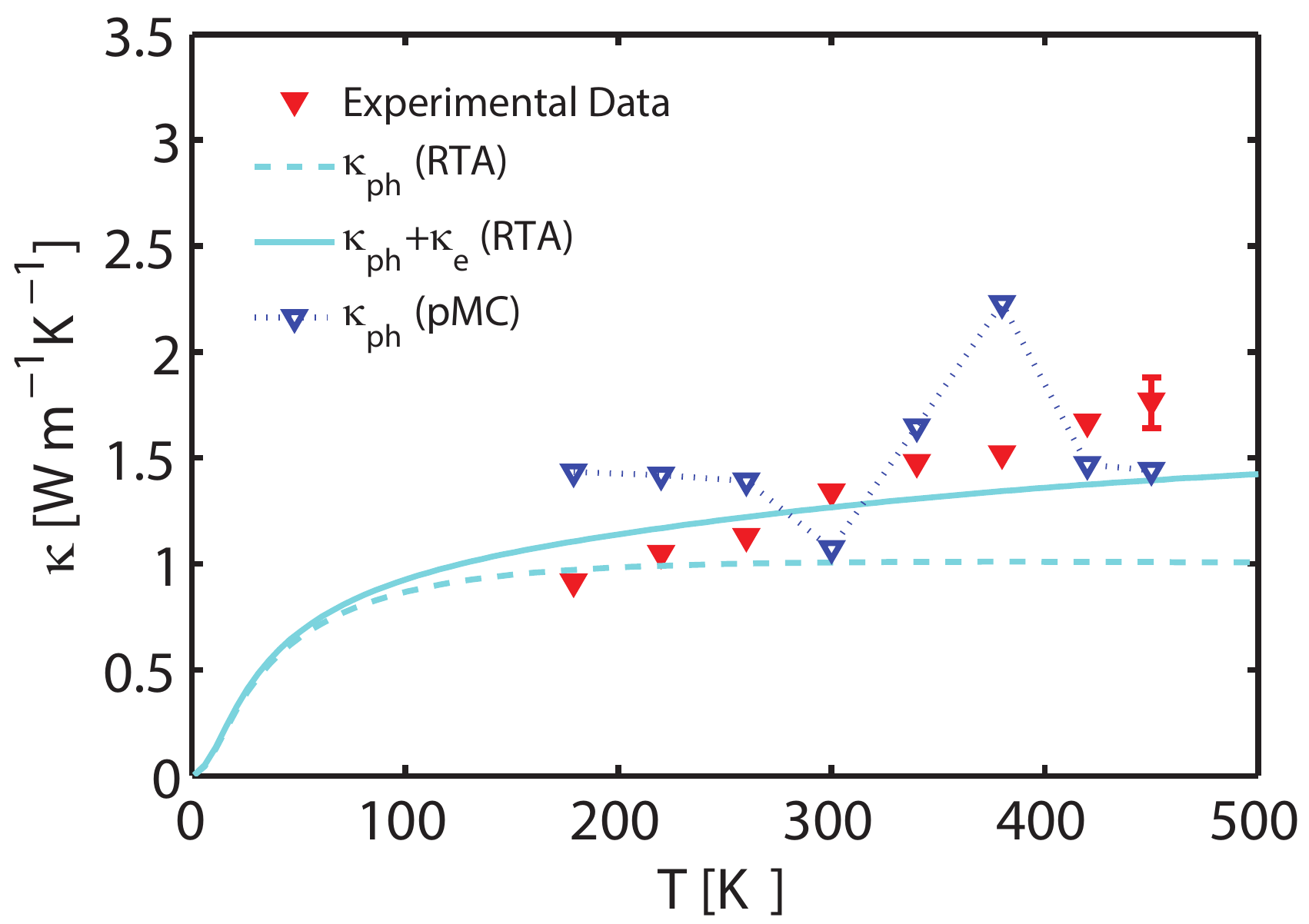}}
	\subfloat[]{\label{fig:kappa27}%
		\includegraphics[width=0.25\textwidth]{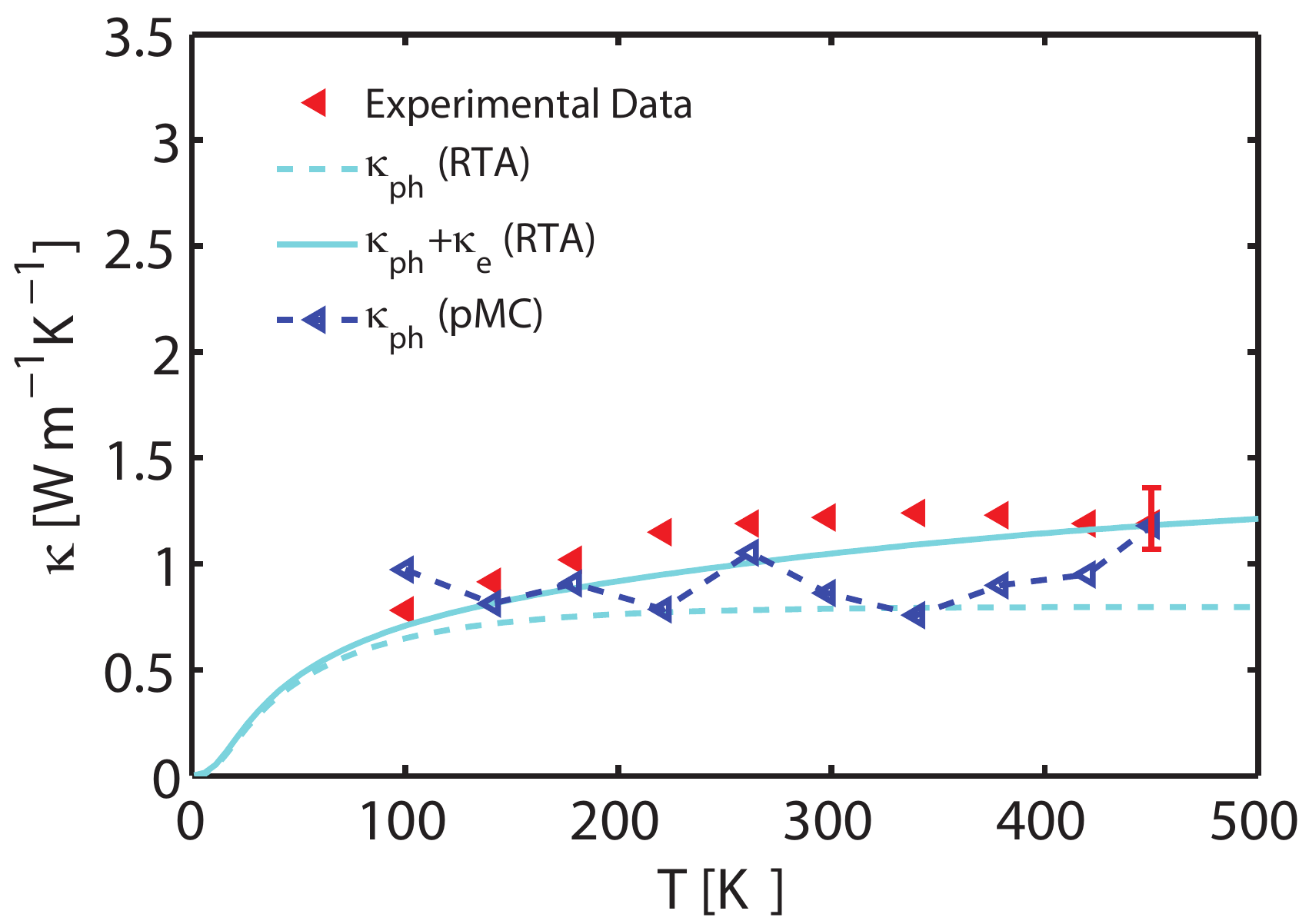}}
	\subfloat[]{\label{fig:kappa41}%
		\includegraphics[width=0.25\textwidth]{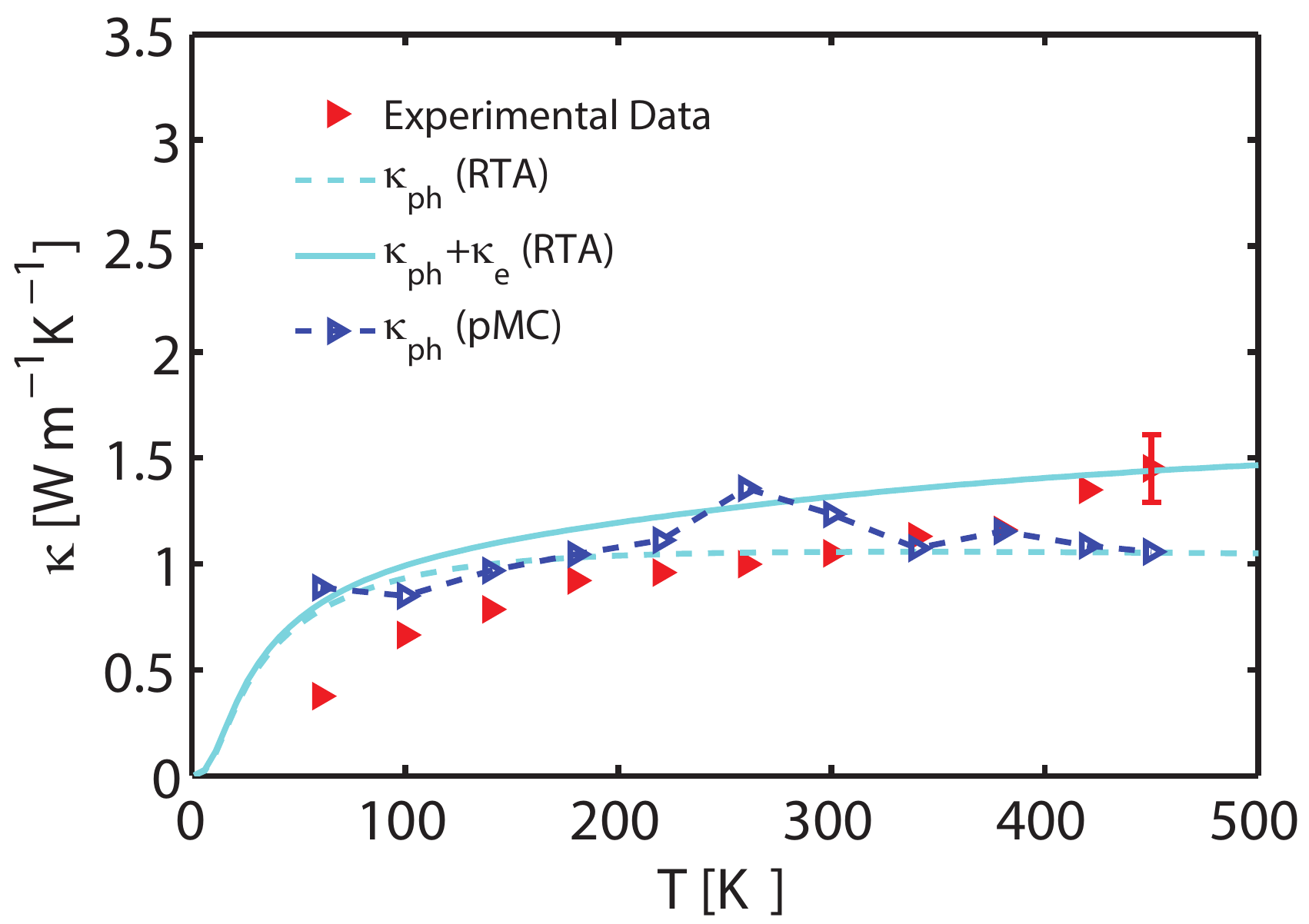}}
	\subfloat[]{\label{fig:kappa86}%
		\includegraphics[width=0.25\textwidth]{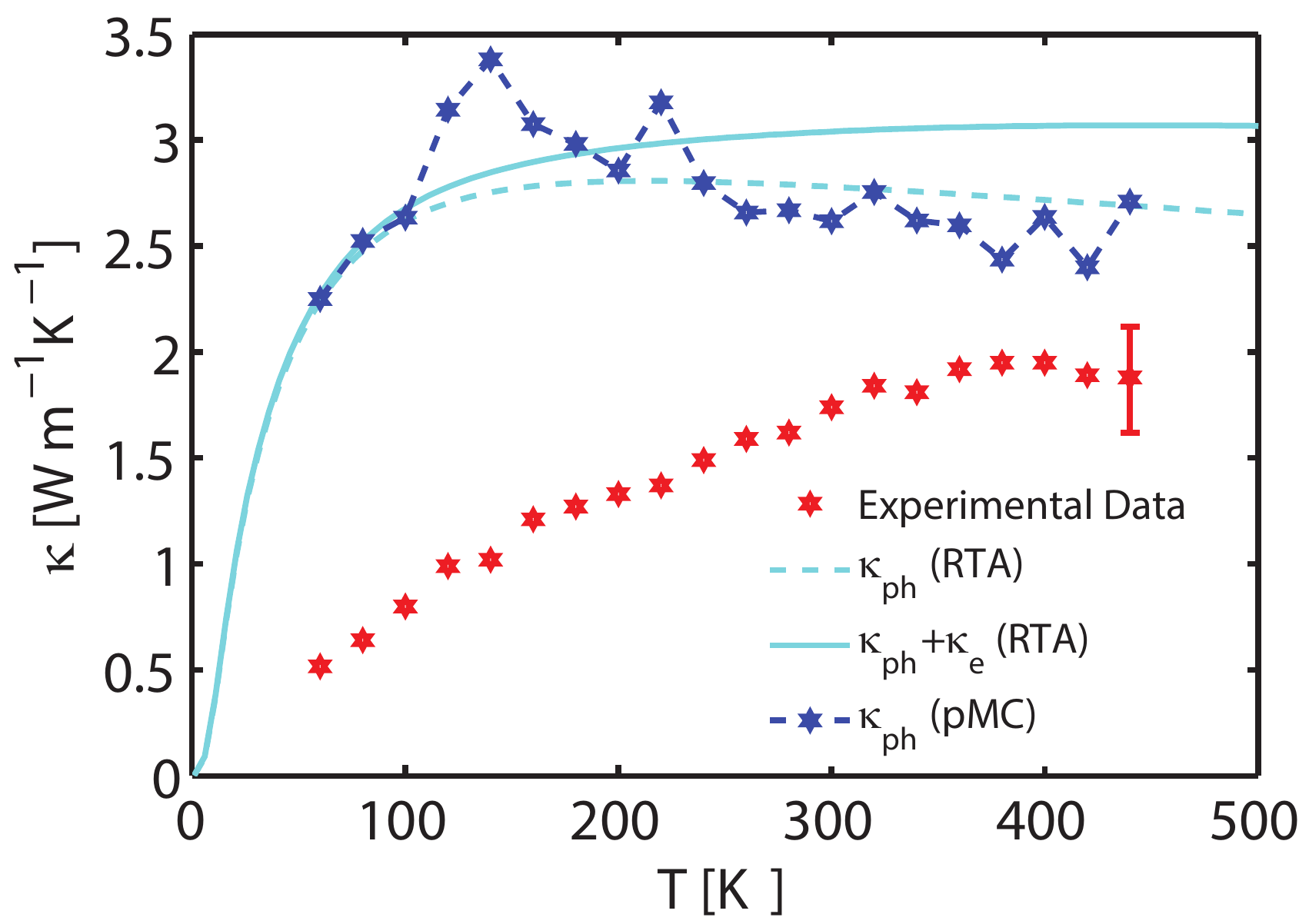}}
	\caption{Thermal conductivity vs. temperature for different wire configurations (a) Length=5.7 $\mu$m, Diameter=56 nm and 6\% Ge (b) Length=6.3 $\mu$m, Diameter=97 nm and 8\% Ge (c) Length=5.6 $\mu$m, Diameter=45 nm and 10\% Ge (d) Length=5 $\mu$m, Diameter=62 nm and 19\% Ge (e) Length=11.6 $\mu$m, Diameter=26 nm and 26\% Ge (f) Length=5.3 $\mu$m, Diameter=26 nm and 27\% Ge (g) Length=5.2 $\mu$m, Diameter=65 nm and 41\% Ge (h) Length=6.2 $\mu$m, Diameter=161 nm and 86\% Ge showing a comparison of results computed based on our RTA (Cyan line) and MC (Blue $-\circ-$) model to experimental values (Red $\bullet$) reported in Ref.~\onlinecite{LeeNL12}.}\label{fig:kvsT}
\end{figure*}

Fig.~\ref{fig:kvsT} shows a comparison of our calculated results to the experimentally measured thermal conductivities reported in Ref.~\onlinecite{LeeNL12} for wire diameters below 100 nm and alloy composition ranging from 6\% Ge (Fig.~\ref{fig:kvsT}a) to 86\% Ge in (Fig.~\ref{fig:kvsT}h). The Monte Carlo simulation results are shown in dashed lines and symbols while the solid lines represent the thermal conductivity from the solution of the Boltzmann transport equation (BTE) in the relaxation time approximation (RTA) including boundary scattering. In Fig.~\ref{fig:kvsT} $\kappa_e$ is the electronic and $\kappa_{ph}$ is the phononic contribution to thermal conductivity. The RTA model has been validated in our previous work \cite{AksamijaPRB13,UpadhyayaJMR15} and the RTA and MC models are in close agreement. Also, our results reproduce the experimental values closely across a wide range of temperatures, diameters, and compositions, with some discrepancy at the highest Ge composition, which may be attributed to the presence of contact resistance, not included in our model.

\subsection{Diameter and roughness dependence}

\begin{figure}[h]
	\centering
	\subfloat{\label{fig:kvsdia}{\subfigimg[width=.85\columnwidth]{(a)}{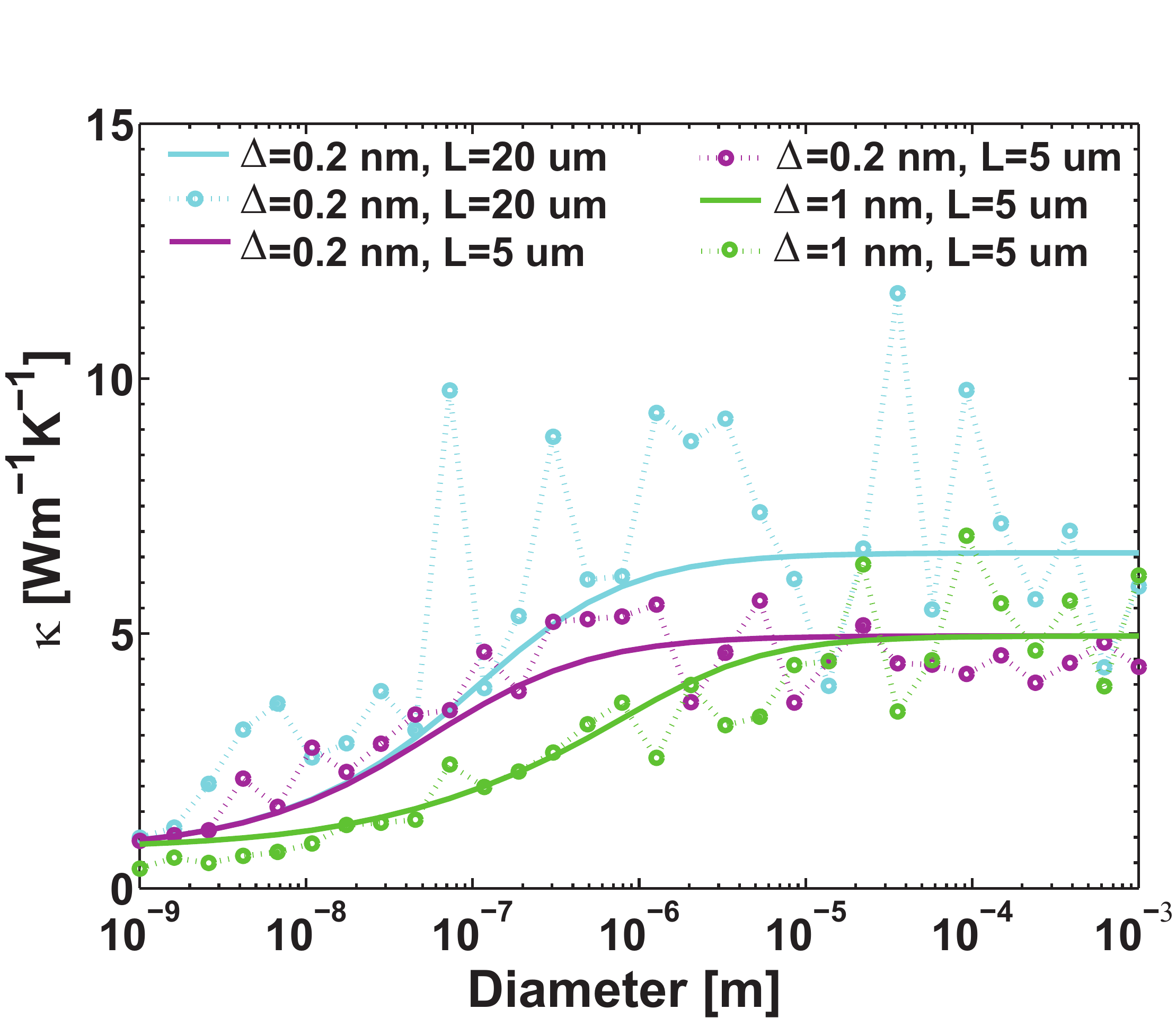}}}\
	\subfloat{\label{fig:kvsSR}{\subfigimg[width=.85\columnwidth]{(b)}{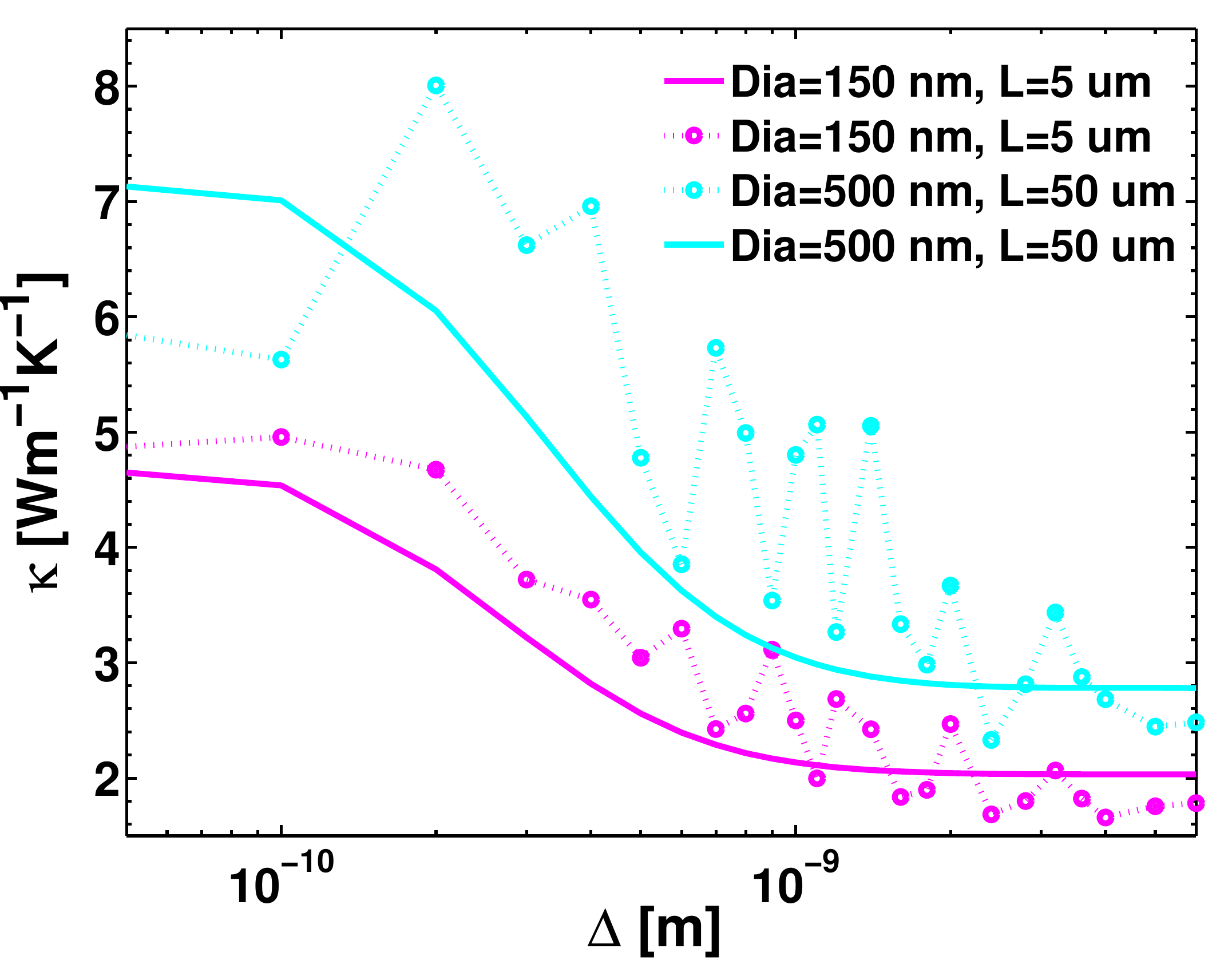}}}
	\caption{(a) Thermal conductivity is plotted as a function of the wire diameter. The diameter dependence is weak due to strong intrinsic scattering. Solid lines represent the BTE results and dashed lines represent Monte Carlo results. (b) Thermal conductivity is plotted as a function of surface roughness. For $\Delta$ values of 1 $\AA$ or less, the conductivity is unaffected and it steadily decreases as the roughness is increased, without any further significant decrease beyond 1 nm.}\label{fig:kvsdiaSR}
\end{figure}

Fig.~\ref{fig:kvsdia} depicts the thermal conductivity vs. diameter for NWs of different lengths and surface roughness $\Delta$. The results were computed for NWs with 20\% Ge concentration at room temperature. The conductivity shows an almost linear diameter dependence at intermediate diameter values for pure Si \cite{XieAPL14}, where boundary scattering is dominant. The deviation from this linear dependence increases with alloying. Alloy scattering rate follows a Rayleigh-like trend ($\tau_M^{-1} \propto \omega^4$ because $g(\omega) \propto \omega^2$ in the long-wavelength regime) and suppresses most of the higher frequency phonons, whereas the low-frequency (long-wavelength) phonons remain nearly unaffected. 

The high-frequency phonons tend to undergo a more diffuse scattering at the boundaries causing a stronger diameter dependence seen in SiNWs, whereas the low frequency phonons undergo a more specular boundary scattering and have smaller diameter dependencies, indicating that it is the mid-range phonons that cause the weak diameter dependence in SiGe NWs. Surface roughness dependence of thermal conductivity is shown in Fig.~\ref{fig:kvsSR}. There is a steady decrease in conductivity with an increase in surface roughness up to 1 nm due to more diffuse boundary scattering, but further increase in roughness does not further reduce the conductivity, which saturates to a value much lower than in pure SiNWs. However the reduction in conductivity due to increased diffuse boundary scattering in SiGe NWs is not as effective as in pure Si NWs, primarily in thin wires where boundary scattering is dominant, due to the supression of the high frequency phonons by alloy scattering. For a pure Si NW of 10 nm diameter and 100 nm length, we calculated a 75\% reduction in conductivity when surface roughness was increased from 0 to 1 nm, whereas the reduction is about 52\% for a SiGe NW of the same dimension.

\subsection{Length dependence}

\begin{figure}[h]
	\centering
	\subfloat{\label{fig:kvsl}{\subfigimg[width=.85\columnwidth]{(a)}{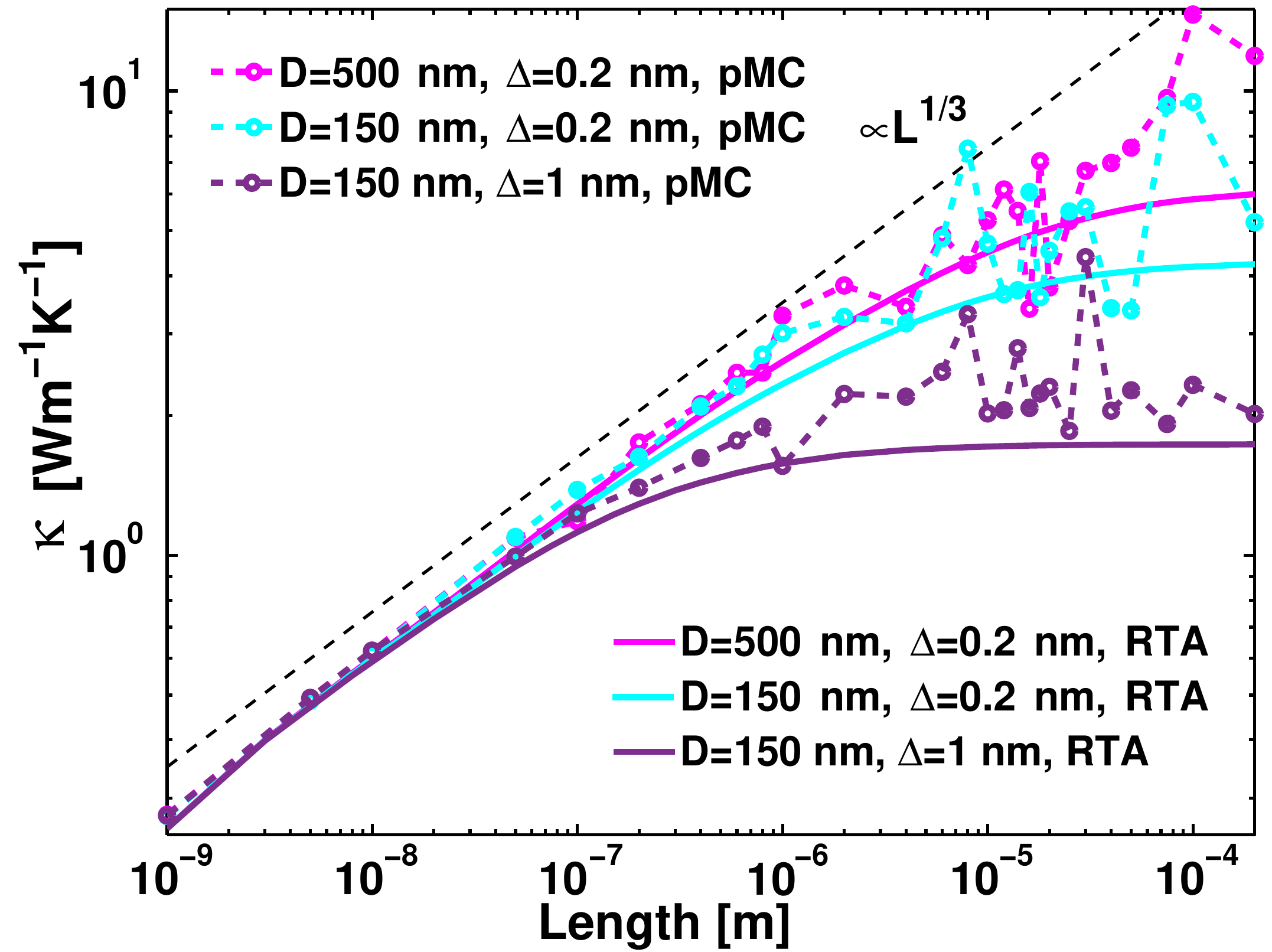}}}\
	\subfloat{\label{fig:alpha}{\subfigimg[width=.85\columnwidth]{(b)}{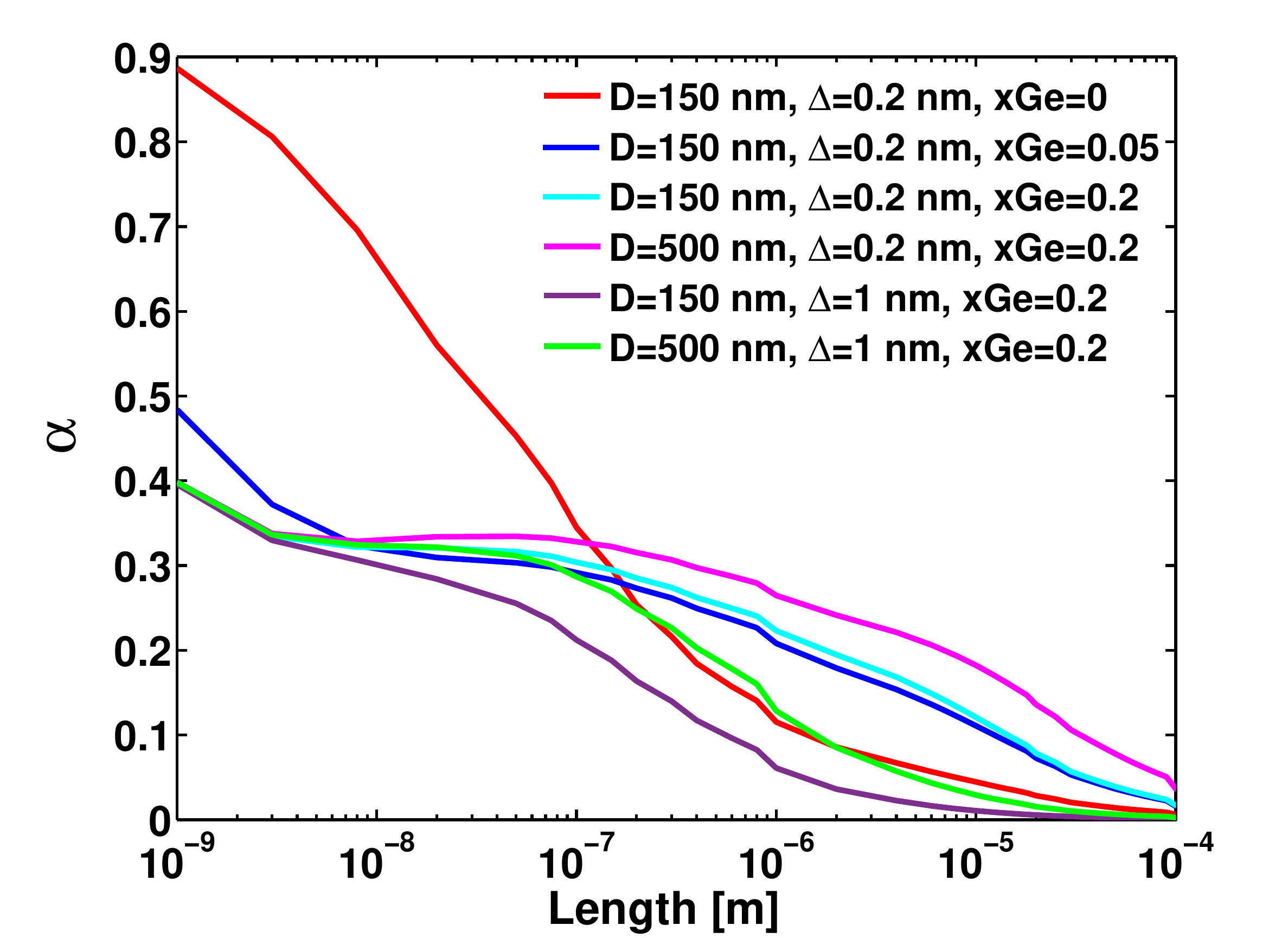}}}
	\caption{(a) Thermal conductivity as a function of length. The solid lines represent the relaxation time approximation (RTA) and dashed lines represent the MC results. The conductivity exhibits a $L^{1/3}$ behavior and gradually transitions into the diffusive regime at lengths exceeding 10 $\mu$m. (b) Exponent of length dependence $\alpha$ as a function of length. In SiGe NWs, $0.3 < \alpha < 0.4$ over a broad range of lengths, indicating non-diffusive transport.}
\end{figure}

Next, we study the dependence of thermal conductivity on the length and composition of the NW. Fig.~\ref{fig:kvsl} shows the dependence of thermal conductivity from pMC simulations on NW length. The dashed lines and symbols are the Monte Carlo results, while solid lines are the deterministic solution of the pBTE in the RTA, shown for comparison, and they are in close agreement. We observe a gradual change in thermal conductivity with length, as seen in Fig.~\ref{fig:kvsl}, with the crossover to purely diffusive transport only occurring at lengths exceeding 10 $\mu$m, far in excess of the average phonon MFP and in agreement with measurements \cite{HsiaoNatNano13}. Alloying suppresses most of the high frequency phonons while the low frequency phonons possess very long MFPs, allowing them to travel several microns without being scattered internally. However, even at very small NW lengths, we do not observe the linear trend in the length dependence that would be characteristic of ballistic transport; instead, we find in Fig.~\ref{fig:kvsl} that the conductivity scales as $L^{1/3}$.  

We plot the running exponent of our results, defined as $\alpha(L)=d \ln\kappa(L)/d \ln L$ \cite{SaitoPRL10}, in Fig.~\ref{fig:alpha} and observe that all SiGe NWs follow the same trend regardless of diameter, with $\alpha<0.4$ even when $L<$10 nm. In contrast, short Si NWs reach the fully ballistic regime (characterized by $\alpha$=1). The $\alpha\approx0.33$ behavior has been observed in many momentum-conserving systems \cite{DharAdvPhys08}, including 1-dimensional chains \cite{LepriPhysRep03,MaiPRL07}, alloy thin films \cite{VermeerschAPL16}, and, over a much narrower range of lengths, even thin Si NWs \cite{YangNT10}. However, the upper limit of length at which we still observe exponent $\alpha\approx1/3$ depends on boundary scattering: in a rough wire ($\Delta$=1 nm), the exponent reduces to the diffusive $\alpha$=0 at a shorter length than in a smooth wire (where $\Delta$=0.2 nm). Diffuse boundary scattering limits the longest MFP and thus results in a more uniform MFP distribution. Hence boundary scattering affects the range of length over which we observe $\alpha=1/3$ but not the length-scaling exponent $\alpha$. Alloy scattering, on the other hand, results in an intrinsically different mode of transport having a broader range of MFPs with very few purely ballistic phonons.

\subsection{Ballistic and diffusive contributions}

\begin{figure}[tbh!]
	\centering
	\subfloat{\label{fig:Gvsl}{\subfigimg[width=.85\columnwidth]{(a)}{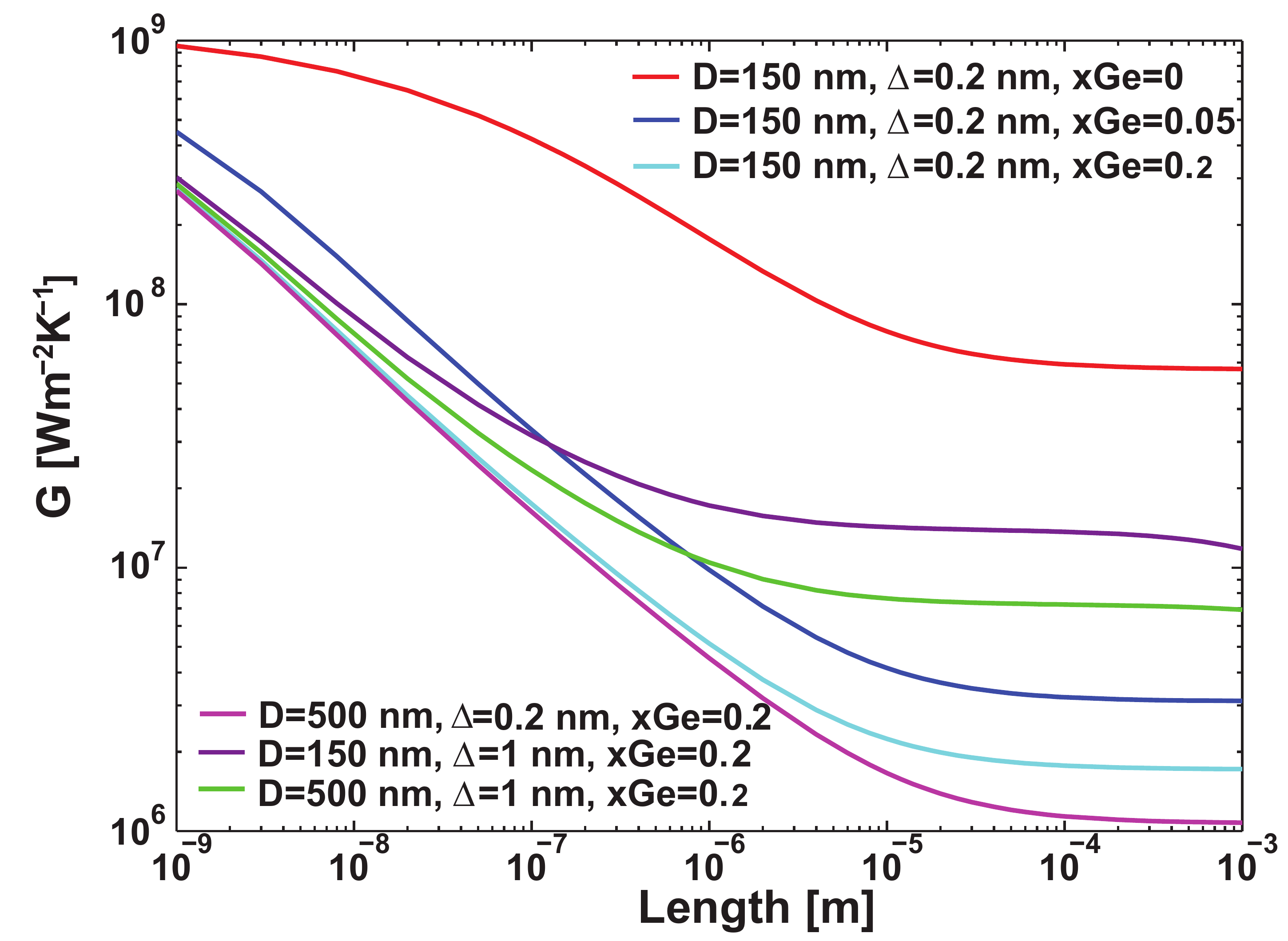}}}\
	\subfloat{\label{fig:kvslambdanew}{\subfigimg[width=.85\columnwidth]{(b)}{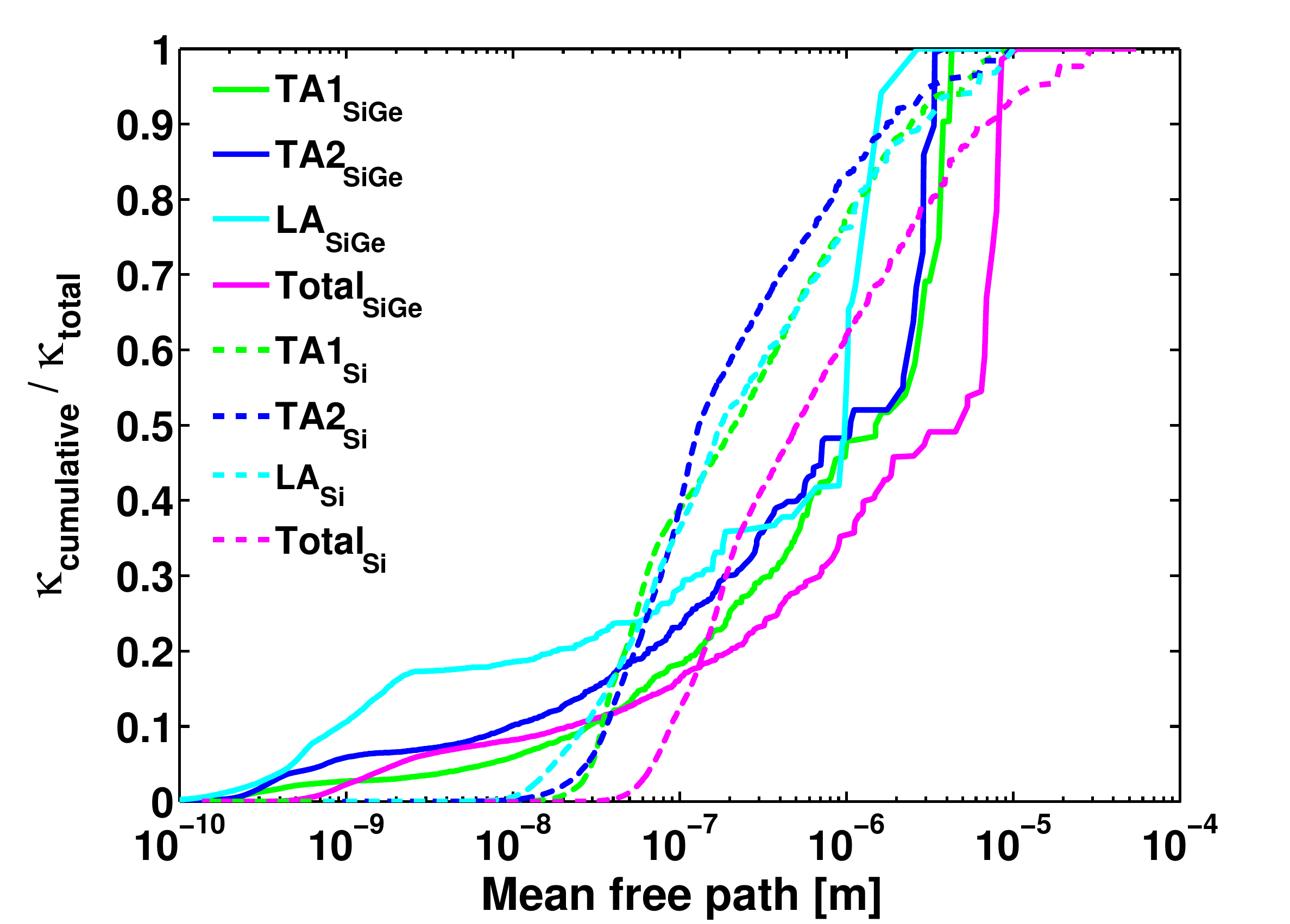}}}
	\caption{(a) Thermal conductance G, as a function of length in Si and alloy NWs of varying diameter, roughness, and composition. (b) Cumulative thermal conductivity as a function of the phonon mean free path for the three acoustic branches and their sum in bulk Si (dashed lines) and Si$_{0.5}$Ge$_{0.5}$ alloy (solid lines), showing the broad range of MFPs in SiGe alloy. Alloys have a broad distribution of MFPs contributing to thermal conduction, with phonons having MFPs in excess of 1 $\mu$m contributing approximately half of the thermal conductivity.}
\end{figure}

In order to shed further light on the origin of the $\kappa(L)\propto L^{1/3}$ length dependence, we first compare it to the ballistic-to-diffusive transition picture suggested by the linear trend in the experimental data \cite{HsiaoNatNano13}: we take the total resistance in the wire as a sum of the ballistic ($G_{Bal}^{-1}$) and diffusive ($L/\kappa_{diff}$) resistances \cite{BaeNC13}, which results in 
\begin{equation} \label{eq:cond}
\kappa(L)=\left(\frac{1}{L G_{Bal}} + \frac{1}{\kappa_{diff}}\right)^{-1} 
\end{equation}
\noindent where, $\kappa(L)$ is the length dependent conductivity, $L$ is the length of the NW, and $\kappa_{diff}$ is the conductivity in the diffusive ($L\rightarrow\infty$) limit. Since $\kappa(L)$ and $\kappa_{diff}$ are known (Fig.~\ref{fig:kvsl}), we can rearrange Eq. \ref{eq:cond} to calculate the ballistic conductance $G_{Bal}$ for different NW lengths. The ballistic conductance $G_{Bal}$ of Si is equal to $10^{9}$ Wm$^{-1}$K$^{-1}$ \cite{MaurerAPL15}. As seen in Fig.~\ref{fig:Gvsl}, when wire length is below 10 nm (much smaller than the MFP in bulk Si) the ballistic conductance of Si NWs plateaus at $10^9$ Wm$^{-1}$K$^{-1}$, matching the theoretical value; in contrast, the conductance in ultrashort SiGe NWs is only $\approx 2\times10^8$ Wm$^{-1}$K$^{-1}$, about 20\% of the ballistic conductance. We also observe that no more than a small fraction of the ballistic conductance is present in SiGe NWs at any length, regardless of diameter and roughness. Hence, transport in alloy NWs is never predominantly ballistic, indicating that the direct ballistic-to-diffusive crossover picture is incomplete. The fraction of ballistic conductance reduces as length increases but about 1\% is still present at lengths exceeding 10 $\mu$m.

The unique properties of alloy nanostructures can be partly analyzed through the prism of the conductivity vs. MFP $\lambda$ plot, shown in Fig.~\ref{fig:kvslambdanew}. We observe a much broader range of MFPs contributing to transport in alloys than in pure Si. There is a large relative contribution to thermal conductivity made by phonons having large MFPs, primarily found in the low energy range of the acoustic phonon branches where both mass disorder (alloy) and anharmonic scattering rates are low, while the boundary scattering is more specular, owing to the large wavelength (small q) of phonons in this range. We conclude that in the Si-Ge alloy, most phonons have very short MFPs and conseqeuntly they make a relatively small contribution to thermal conductivity, while fewer phonons have very long MFPs exceeding one micrometer, but they make a substantial contribution to thermal transport. The 50$\%$ accumulation point where one half of the total thermal conductivity is reached corresponds to MFPs of around one micrometer, implying that half of the heat is carried by phonons with MFPs exceeding a micron; in contrast, such phonons only contribute around 20$\%$ in pure Si. 

\begin{figure*}[tb]
	\subfloat{\label{fig:sidis}{\subfigimg[width=0.4\textwidth]{(a)}{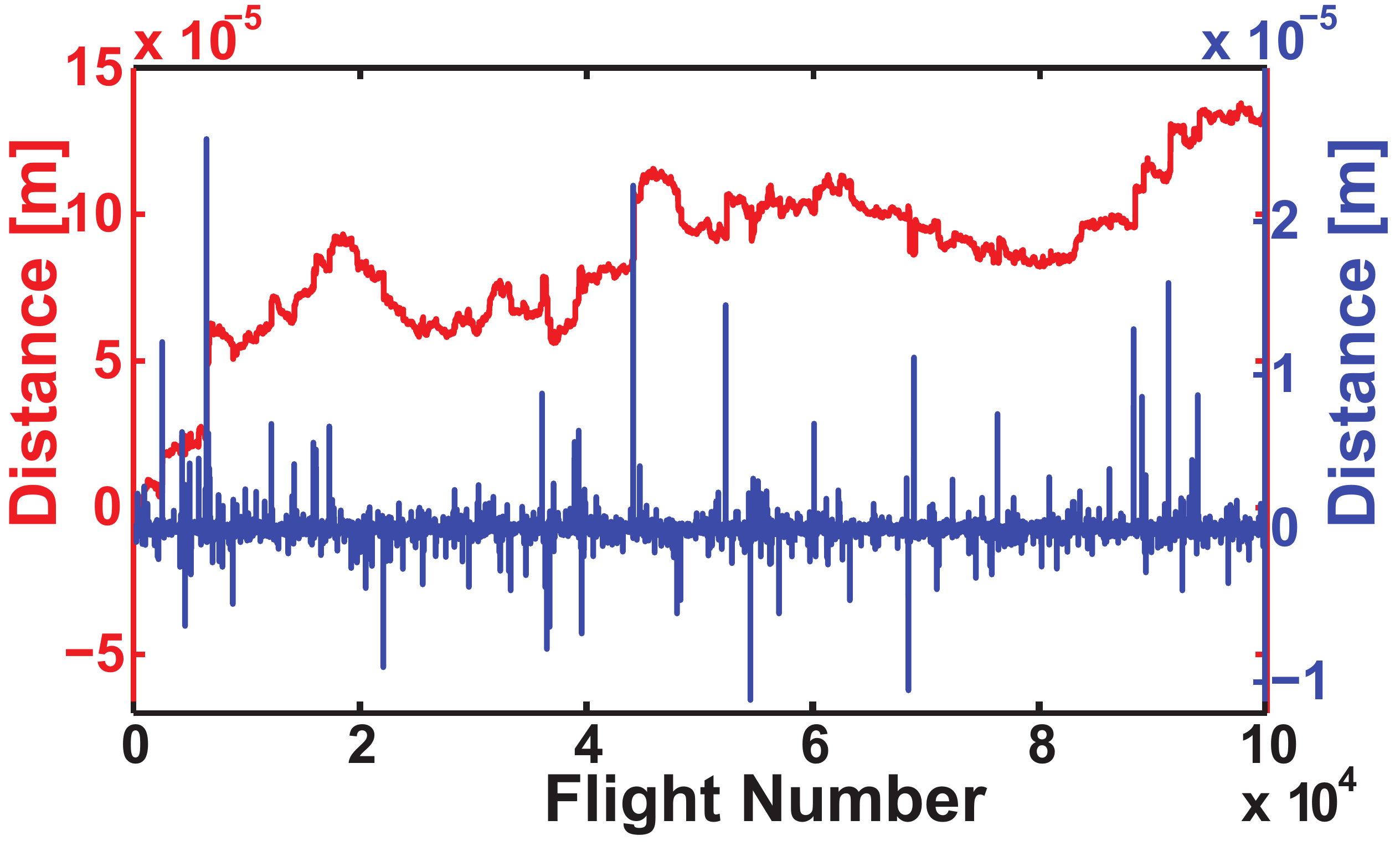}}}%
	\subfloat{\label{fig:sitime}{\subfigimg[width=0.4\textwidth]{(b)}{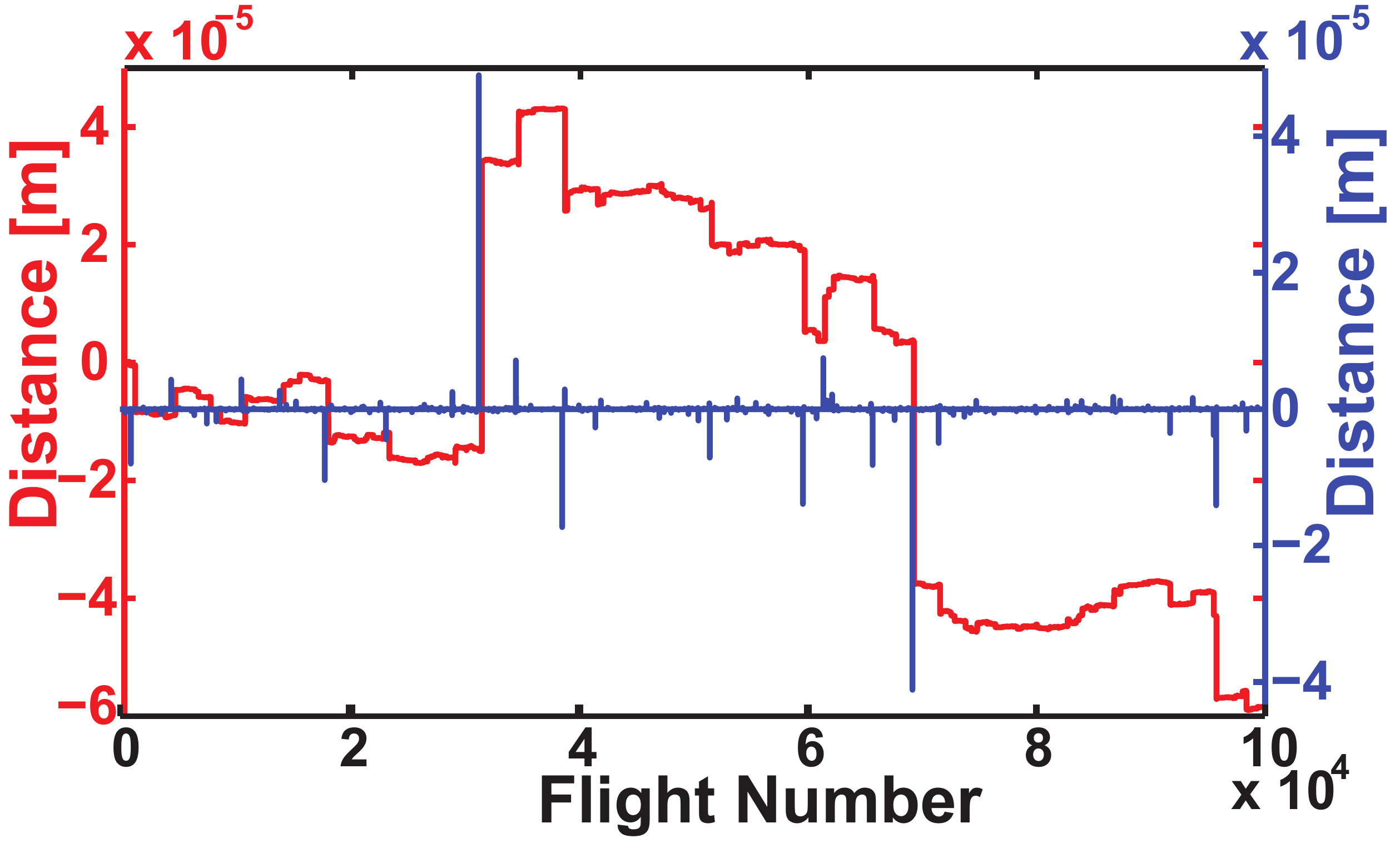}}}\
	\subfloat{\label{fig:sigedis}{\subfigimg[width=0.4\textwidth]{(c)}{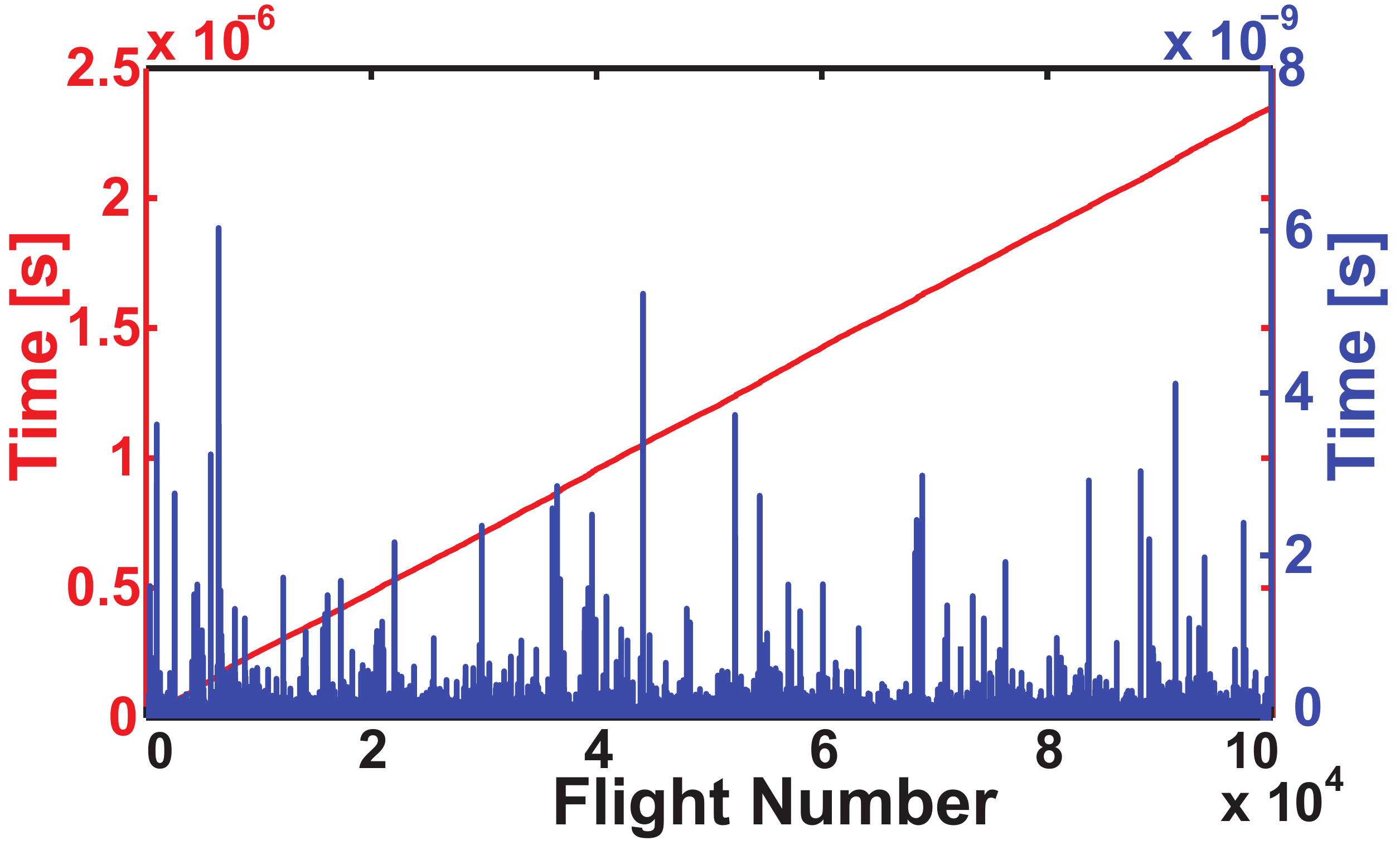}}}%
	\subfloat{\label{fig:sigetime}{\subfigimg[width=0.4\textwidth]{(d)}{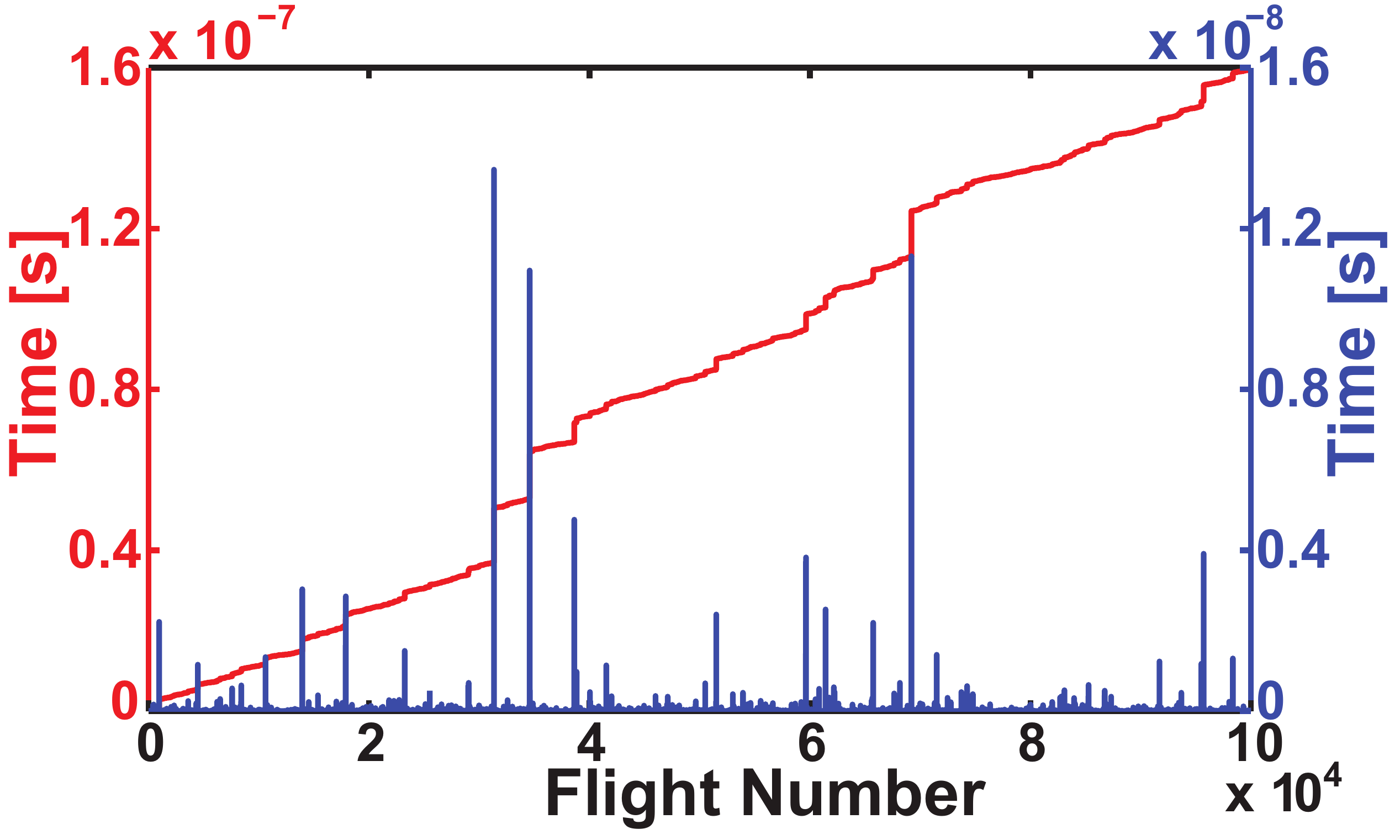}}}
	\caption{The distance traveled by one simulated phonon between successive scattering events (blue dotted line), and the cumulative distance (red line) vs. flight number is plotted for (a) a pure Si nanowire and (b) a Si$_{0.8}$Ge$_{0.2}$ nanowire.  Also, the time between scattering events (blue dotted line) and the cumulative time (red line) is plotted vs. the flight number in (c) pure Si NW and (d) Si$_{0.8}$Ge$_{0.2}$ NW. 
	}\label{fig:dis&time}
\end{figure*}
\subsection{Phonon flight length}
Next, we plot the distance along the NW traveled by one simulated phonon through a sequence of free flights interrupted by scattering events. Each flight number in the plot corresponds to a single free-flight between successive scattering events for both Si (Fig.~\ref{fig:sidis}) and Si$_{0.8}$Ge$_{0.2}$ (Fig.~\ref{fig:sigedis}). We focus on a Ge composition of 0.2 as that is typically found to be optimal for TE applications \cite{ViningJAP91_2}. One phonon is chosen from the ensemble at random as a representative and plotted, having found all of the simulated phonons to exhibit qualitatively similar behavior. The scattering is predominantly elastic, arising in both cases from boundary roughness, while in alloy NWs there is a strong additional component due to mass disorder. It is interspersed by less frequent inelastic (anharmonic phonon-phonon) events which mix the phonon modes. We find Si NWs to have a more continuous distribution of distances due to the series of relatively uniform flights characteristic of diffusive transport, leading to a white-noise-like appearance seen in Fig.~\ref{fig:sidis}, which is readily associated with Brownian motion and predominantly diffusive transport.

In contrast, phonon flights in SiGe NWs are comprised of sequences of many short flights interrupted by rare long leaps, as evidenced by the micron-sized jumps in the distance traveled by the phonon shown in Fig.~\ref{fig:sigedis}. This behavior in the alloy is a consequence of the strong mass disorder scattering, which has a Rayleigh-like dependence on phonon frequency ($\tau_M^{-1}\propto \omega^4$) and affects the upper portion of the phonon spectrum far more than the low-frequency modes. Viewing the whole phonon ensemble collectively, the result is a heavy-tailed distribution of free-flight lengths, shown in Fig.~\ref{fig:histogram}. The difference is especially prominent when we compare alloy NWs to pure Si NWs, in which the tail of the phonon flight distribution decays faster. Phonons making long jumps exceeding 1$\mu$m are more than twice as frequent in SiGe NWs as they are in Si NWs, as seen in the distribution of flight lengths $\Delta x$ as measured in the direction of heat flow along the NW. The difference alloy and non-alloy steadily increases for longer leaps, in spite of alloy scattering causing phonons in SiGe to have MFPs more than an order of magnitude shorter on average than pure Si. The heavy-tailed behavior is characteristic of L\'{e}vy walk dynamics \cite{DharPRE13,ZaburdaevRMP15}, which has already been linked to superdiffusive phonon transport in low-dimensional \cite{CiprianiPRL05} and alloy systems \cite{VermeerschPRB15}. 

\subsection{Diffusion coefficient}

\begin{figure}[tb]
	\centering
	\subfloat{\label{fig:histogram}{\subfigimg[width=.85\columnwidth]{(a)\quad}{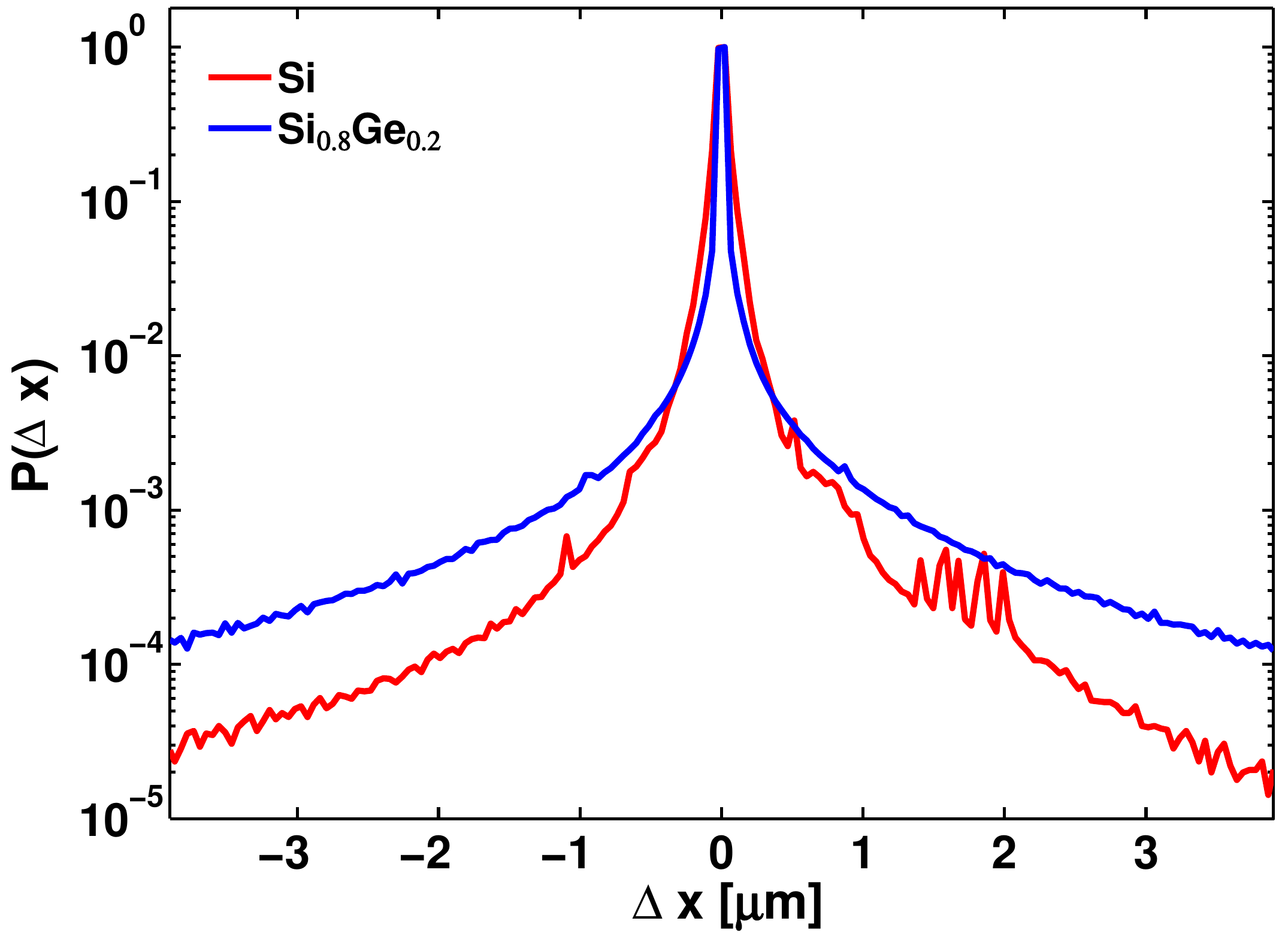}}}\
	\subfloat{\label{fig:msd}{\subfigimg[width=.85\columnwidth]{(b)\quad}{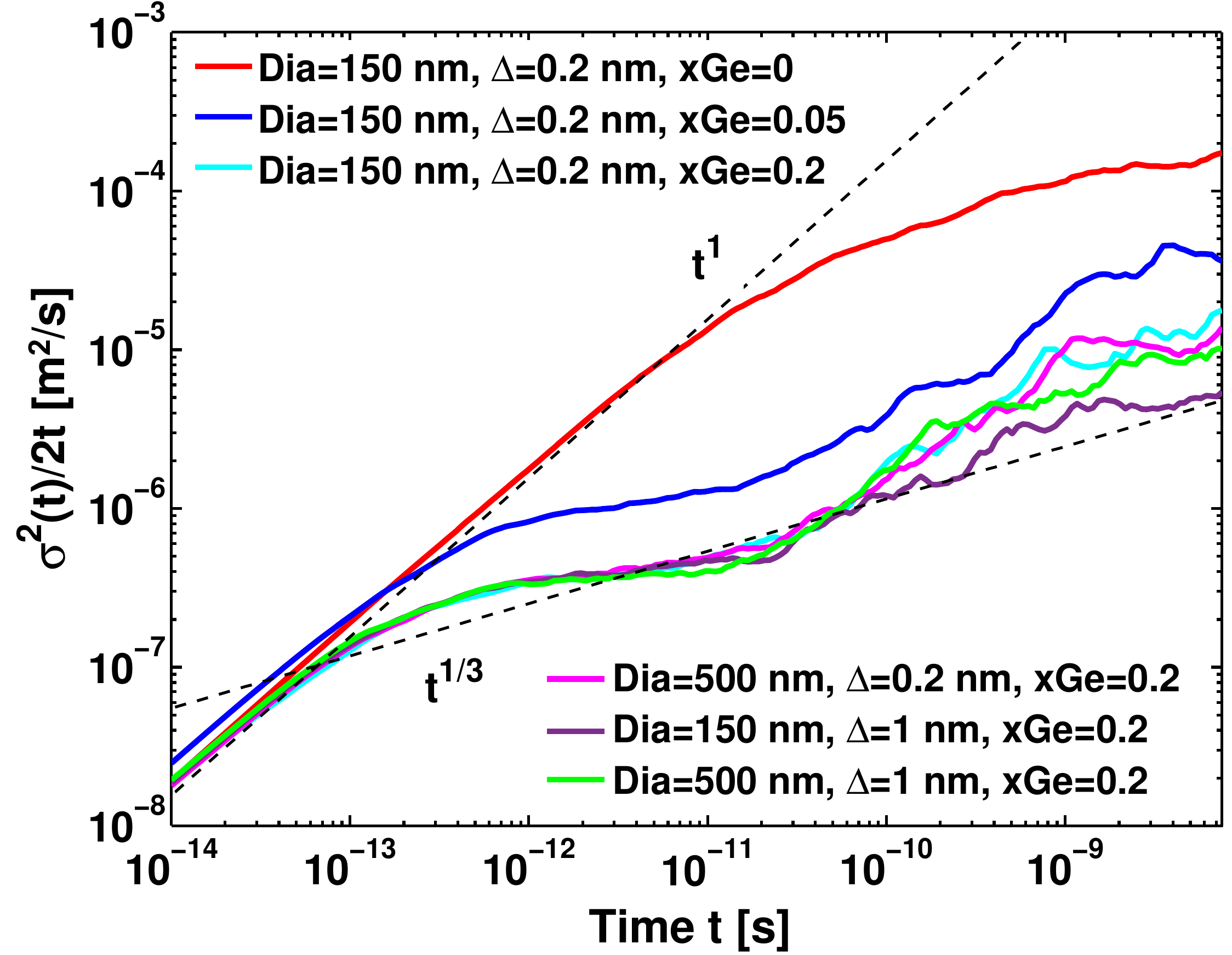}}}
	\caption {(a) Histogram of the individual free-flight lengths in Si (red) and Si$_{0.8}$Ge$_{0.2}$ NWs (blue). SiGe NWs show a larger proportion of long leaps, which leads to a heavy-tailed distribution. (b) diffusion coefficient vs. time. In SiGeNWs we observe a broad intermediate regime in which the exponent of the diffusion coefficient is $\approx 0.33$ over several orders of magnitude in time, whereas in SiNWs we observe $\alpha=1$ in the ballistic regime followed by a smooth transition into the diffusive ($\alpha=0$) regime.}\label{fig:alpha&msd}
\end{figure}

We further study the non-diffusive behaviour of phonons in alloy NWs by examining the time-dependent phonon transport calculated from our pMC simulations. The time dependence of the mean square energy displacement (MSD)\cite{DenisovPRL03} is calculated from $\sigma^2(t)=\left<\Delta x^2(t)\right>$ and related to an exponent $\sigma^2(t)\propto t^\beta$. Consequently, the diffusion coefficient $\sigma^2(t)/2t \propto t^{\beta-1}$. The exponent of length dependence $\alpha$, shown previously in Fig.~\ref{fig:alpha}, has also been related to the MSD through $\alpha=\beta-1$ \cite{LiuPRL14}. In normal diffusion, where Fourier's law remains valid, phonons undergo Brownian motion resulting in $\beta=1$ (as $\sigma^2(t)\propto t$) \cite{ShlesingerNat93}, which also means the exponent of length dependence $\alpha=0$ and the conductivity is constant, independent of length. In contrast, when the system size is smaller than the mean free path, phonon flights are uninterrupted by scattering and their distance from origin grows in proportion with time; hence, ballistic MSD is quadratic in time ($\beta=2$), implying that $\alpha=1$ and the conductivity is linearly proportional to length \cite{LiNJP15}.

Our observed $\alpha=1/3$ scaling is attributed to an intermediate regime of superdiffusion: when $1<\beta<2$ transport is neither entirely ballistic nor diffusive. Instead, it is collectively characterized by a mix of long quasi-ballistic leaps, interrupted by bursts of short, diffusive steps. Consequently, the diffusion coefficient $\sigma^2(t)/2t\propto t^\alpha$ should also imply a length exponent $0<\alpha<1$ in the superdiffusive regime. We plot the diffusion coefficient $\sigma^2(t)/2t$ as a function of simulation time in Fig.~\ref{fig:msd}, and observe that $\alpha=1/3$ (and $\beta=4/3$) over a wide range of time scales in SiGeNWs. In bulk SiGe alloys and SiGe thin films in the cross-plane direction, Vermeersch et al.\cite{VermeerschPRB15,VermeerschAPL16} have shown that $\beta$=1.34, in close agreement with our findings, confirming that our length scaling is a consequence of superdiffusion.

The range of lengths over which we observe superdiffusion ($\alpha \approx 1/3$ over 10 nm$<L<$ 10 $\mu$m) far exceeds the average phonon MFP in the system; instead, it maps directly to the wide range of timescales over which superdiffusion is observed here in SiGeNWs (2 ps$<t<$2 ns in Fig.~\ref{fig:msd}) using the simple cut-off $t_s \approx L/v_s$ for superdiffusion in finite systems\cite {LepriPhysRep03,LiuPRL14}, with $v_s$ being the speed of sound ($v_s \approx$ 5000 m/s in SiGe, depending on alloy composition). We observe that diameter and roughness do not alter the exponent of the diffusion coefficient; instead, they affect only the onset of the transition from superdiffusion ($0<\alpha<1$) into purely diffusive regime ($\alpha=0$), thus reducing the resulting conductivity in the steady-state analogously to its length dependence in Fig.~\ref{fig:kvsl}.

\section{Conclusion}\label{sec:conclusion}

We employ the phonon Monte Carlo method to study thermal transport in SiGe NWs. 
We find that thermal conductivity scales as $L^{1/3}$ over a broad range of lengths and conclude that in SiGe NWs the direct ballistic-to-diffusive crossover picture is incomplete and should be augmented by superdiffusion in the broad intermediate range of NW length from 10 nm to 10 $\mu$m. 
Our study shows that alloy nanostructures exhibit novel heat dynamics and can be used as a unique fundamental platform to study the breakdown of Fourier's law. The superdiffusive transport is brought on by L\'{e}vy-like heavy-tailed distribution of phonon flights, and causes a length dependent thermal conductivity with $\kappa \propto L^{1/3}$ over a broad range of lengths extending from 10 nm all the way to 10 $\mu$m, far exceeding phonon MFP or NW diameter. Thus, lattice conductivity is length-tunable even in NWs several microns long, with potential applications to reducing thermal conductivity and thus increasing thermoelectric figure-of-merit in alloy NWs with sub-ten-micron lengths.

\appendix*
\section{Phonon Scattering Rates}\label{sec:appendix}

We find the combined lifetime with both intrinsic scattering (from anharmonic phonon-phonon and mass difference interactions inside the wire) and partially diffuse boundary scattering at the rough boundary in circular wires. The phonon lifetime $\tau_{int.}(\vec q)$ combines all the intrinsic scattering mechanisms, including anharmonic 3-phonon interactions, impurity, isotope, and alloy mass-difference scattering. The resistive umklapp phonon-phonon scattering rate is calculated in the general approximation for dielectric crystals \cite{MorelliPRB02,UpadhyayaJMR15}
\begin{equation} \label{eq:rateU}
\tau^{-1}_{\lambda,\mathrm{U}}(\vec q)=\frac{\hbar \gamma_{\lambda}^2}{\overline{M} \Theta_{\lambda} \bar{\upsilon}_\lambda^2}\omega^2_{\lambda}(\vec q) T e^{-\Theta_{\lambda}/3T},
\end{equation}
\noindent where the speed of sound $\bar{\upsilon}_\lambda$ of each branch $\lambda$ is determined from the average slope of its dispersion curve near the $\Gamma$ point \cite{KongPRB09,KlemensCarbon94} and $\overline{M}$ is the average atomic mass. The exponential term $e^{-\Theta_{\lambda}/3T}$ in the temperature dependence controls the onset of resistive umklapp scattering for each phonon branch through the branch-specific Debye temperatures $\Theta_{\lambda}$, which were obtained from \cite{SlackSSPv34}
\begin{equation} \label{eq:debye}
\Theta_{\lambda}^2=\frac{5\hbar^2}{3k_B^2}\frac{\int \omega^2 g_{\lambda}(\omega) d\omega}{\int g_{\lambda}(\omega)d\omega},
\end{equation}
\noindent where the vibrational density of states (vDOS) function $g_{\lambda}(\omega)=\sum_{\vec q}\delta\left[\omega-\omega_{\lambda}(\vec q)\right]$ was calculated for each phonon branch $\lambda$ from the full dispersion. 

Scattering from mass differences due to the presence of mass variation in the alloy can be represented by an energy-dependent rate 
\begin{equation}
\tau^{-1}_{\mathrm{M}}(\omega) = \frac{\pi}{6} \Omega_0 \Gamma \omega^2 g(\omega), 
\end{equation}
\noindent with the total vDOS function given by a sum over all branches $g(\omega)=\sum_{\lambda}g_{\lambda}(\omega)$ \cite{GargPRL11}. The mass-difference constant $\Gamma$ is given by the sum over all the participating elements weighted by their mass $M_i$ relative to the average mass $\overline{M}$ \cite{WeiPRL93}, $\Gamma=\sum_i f_i \left(1-M_i/\overline{M}\right)^2$.

Boundary roughness scattering is characterized through a momentum-dependent specularity parameter \cite{SofferJAP67} $p(\vec q)=\exp(-\langle \phi^2\rangle)$, where $\phi(\vec q,\vec r)=2\vec q\cdot{\hat s}z(\vec r)=2q z(\vec r) \cos{\Theta_B}$ is the phase difference between the incoming wave and the outgoing specularly reflected wave at point $\vec r$ where the surface normal unit vector is $\hat s$. We assume that the surface height $z(\vec r)$ is a random function of position on the rough boundary $\vec r$ with a Gaussian distribution, so that $\langle z\rangle=0$ and $\langle z^2\rangle=\Delta^2$, where $\Delta$ is the rms height of the surface roughness \cite{GoodnickPRB85}.


\begin{thebibliography}{10}
	
	\bibitem{SnyderNMAT08}
	G.~Jeffrey Snyder and Eric~S. Toberer.
	\newblock Complex thermoelectric materials.
	\newblock {\em Nature Mater.}, 7:105--114, 2008.
	
	\bibitem{HochbaumNAT08}
	A.I. Hochbaum, R.~Chen, R.D. Delgado, W.~Liang, E.C. Garnett, M.~Najarian,
	A.~Majumdar, and P.~Yang.
	\newblock Enhanced thermoelectric performance of rough silicon nanowires.
	\newblock {\em Nature}, 451:163, 2008.
	
	\bibitem{MartinPRL09}
	Pierre Martin, Zlatan Aksamija, Eric Pop, and Umberto Ravaioli.
	\newblock Impact of phonon-surface roughness scattering on thermal conductivity
	of thin si nanowires.
	\newblock {\em Phys. Rev. Lett.}, 102(12):125503, 2009.
	
	\bibitem{ShiAPL09}
	Lihong Shi, Donglai Yao, Gang Zhang, and Baowen Li.
	\newblock Size dependent thermoelectric properties of silicon nanowires.
	\newblock {\em Appl. Phys. Lett.}, 95(6):063102, 2009.
	
	\bibitem{ChenAPL09}
	Jie Chen, Gang Zhang, and Baowen Li.
	\newblock Tunable thermal conductivity of si[sub 1 - x]ge[sub x] nanowires.
	\newblock {\em Appl. Phys. Lett.}, 95(7):073117, 2009.
	
	\bibitem{KimAPL10}
	Hyoungjoon Kim, Ilsoo Kim, Heon jin Choi, and Woochul Kim.
	\newblock Thermal conductivities of si[sub 1 - x]ge[sub x] nanowires with
	different germanium concentrations and diameters.
	\newblock {\em Appl. Phys. Lett.}, 96:233106, 2010.
	
	\bibitem{YinAPL12}
	Liang Yin, Eun Kyung~Lee, Jong Woon~Lee, Dongmok Whang, Byoung Lyong~Choi, and
	Choongho Yu.
	\newblock The influence of phonon scatterings on the thermal conductivity of
	sige nanowires.
	\newblock {\em Appl. Phys. Lett.}, 101(4):043114, 2012.
	
	\bibitem{LeeNL12}
	Eun~Kyung Lee, Liang Yin, Yongjin Lee, Jong~Woon Lee, Sang~Jin Lee, Junho Lee,
	Seung~Nam Cha, Dongmok Whang, Gyeong~S. Hwang, Kedar Hippalgaonkar, Arun
	Majumdar, Choongho Yu, Byoung~Lyong Choi, Jong~Min Kim, and Kinam Kim.
	\newblock Large thermoelectric figure-of-merits from sige nanowires by
	simultaneously measuring electrical and thermal transport properties.
	\newblock {\em Nano Lett.}, 12:2918--2923, 2012.
	
	\bibitem{HsiaoNatNano13}
	Tzu-Kan Hsiao, Hsu-Kai Chang, Sz-Chian Liou, Ming-Wen Chu, Si-Chen Lee, and
	Chih-Wei Chang.
	\newblock Observation of room-temperature ballistic thermal conduction
	persisting over 8.3 [micro]m in sige nanowires.
	\newblock {\em Nat Nano}, 8(7):534--538, 2013.
	
	\bibitem{VermeerschPRB15}
	Bjorn Vermeersch, Jes\'us Carrete, Natalio Mingo, and Ali Shakouri.
	\newblock Superdiffusive heat conduction in semiconductor alloys. i.
	theoretical foundations.
	\newblock {\em Phys. Rev. B}, 91:085202, 2015.
	
	\bibitem{BlumenPRA89}
	A.~Blumen, G.~Zumofen, and J.~Klafter.
	\newblock Transport aspects in anomalous diffusion: L\'evy walks.
	\newblock {\em Phys. Rev. A}, 40:3964--3973, 1989.
	
	\bibitem{BarthelemyNature08}
	Pierre Barthelemy, Jacopo Bertolotti, and Diederik~S. Wiersma.
	\newblock A l\'evy flight for light.
	\newblock {\em Nature}, 453:495--498b, 2008.
	
	\bibitem{LiAPL03_1}
	D.~Li, Y.~Wu, P.~Kim, L.~Shi, P.~Yang, and A.~Majumdar.
	\newblock Thermal conductivity of individual silicon nanowires.
	\newblock {\em Appl. Phys. Lett.}, 83:2934--2936, 2003.
	
	\bibitem{XieAPL14}
	Guofeng Xie, Yuan Guo, Xiaolin Wei, Kaiwang Zhang, Lizhong Sun, Jianxin Zhong,
	Gang Zhang, and Yong-Wei Zhang.
	\newblock Phonon mean free path spectrum and thermal conductivity for
	si1âˆ’xgex nanowires.
	\newblock {\em Appl. Phys. Lett.}, 104(23):233901, 2014.
	
	\bibitem{XiePCCP13}
	Guofeng Xie, Yuan Guo, Baohua Li, Liwen Yang, Kaiwang Zhang, Minghua Tang, and
	Gang Zhang.
	\newblock Phonon surface scattering controlled length dependence of thermal
	conductivity of silicon nanowires.
	\newblock {\em Phys. Chem. Chem. Phys.}, 15:14647--14652, 2013.
	
	\bibitem{WangAPL10}
	Zhao Wang and Natalio Mingo.
	\newblock Diameter dependence of sige nanowire thermal conductivity.
	\newblock {\em Appl. Phys. Lett.}, 97(10):101903, 2010.
	
	\bibitem{AksamijaPRB13}
	Z.~Aksamija and I.~Knezevic.
	\newblock Thermal conductivity of si$_{1-x}$ge$_x$/si$_{1-y}$ge$_y$
	superlattices: Competition between interfacial and internal scattering.
	\newblock {\em Phys. Rev. B}, 88:155318, 2013.
	
	\bibitem{SellanJAP10}
	D.~P. Sellan, J.~E. Turney, A.~J.~H. McGaughey, and C.~H. Amon.
	\newblock Cross-plane phonon transport in thin films.
	\newblock {\em J. Appl. Phys.}, 108(11):113524, 2010.
	
	\bibitem{AksamijaPRB10}
	Z.~Aksamija and I.~Knezevic.
	\newblock Anisotropy and boundary scattering in the lattice thermal
	conductivity of silicon nanomembranes.
	\newblock {\em Phys. Rev. B}, 82:045319, 2010.
	
	\bibitem{SofferJAP67}
	Stephen~B. Soffer.
	\newblock Statistical model for the size effect in electrical conduction.
	\newblock {\em J. Appl. Phys.}, 38(4):1710--1715, 1967.
	
	\bibitem{MazumderJHT01}
	S.~Mazumder and A.~Majumdar.
	\newblock Monte carlo study of phonon transport in solid thin films including
	dispersion and polarization.
	\newblock {\em J. Heat Transf.}, 123(4):749, 2001.
	
	\bibitem{ChenJHT05}
	Yunfei Chen, Deyu Li, Jennifer~R. Lukes, and Arun Majumdar.
	\newblock Monte carlo simulation of silicon nanowire thermal conductivity.
	\newblock {\em J. Heat Transf.}, 127(10):1129--1137, 2005.
	
	\bibitem{LacroixAPL06}
	D.~Lacroix, K.~Joulain, D.~Terris, and D.~Lemonnier.
	\newblock Monte carlo simulation of phonon confinement in silicon
	nanostructures: Application to the determination of the thermal conductivity
	of silicon nanowires.
	\newblock {\em Appl. Phys. Lett.}, 89(10):103104, 2006.
	
	\bibitem{JengJHT08}
	Ming-Shan Jeng, Ronggui Yang, David Song, and Gang Chen.
	\newblock Modeling the thermal conductivity and phonon transport in
	nanoparticle composites using monte carlo simulation.
	\newblock {\em J. Heat Transf.}, 130:042410, 2008.
	
	\bibitem{Randrianalisoa08}
	J.~Randrianalisoa and D/~Baillis.
	\newblock Monte carlo simulation of steady--state microscale phonon heat
	transport.
	\newblock {\em J. Heat Transf.}, 130:072404, 2008.
	
	\bibitem{PeraudPRB11}
	Jean-Philippe~M. P\'eraud and Nicolas~G. Hadjiconstantinou.
	\newblock Efficient simulation of multidimensional phonon transport using
	energy-based variance-reduced monte carlo formulations.
	\newblock {\em Phys. Rev. B}, 84:205331, 2011.
	
	\bibitem{RamayyaPRB12}
	E.~B. Ramayya, L.~N. Maurer, A.~H. Davoody, and I.~Knezevic.
	\newblock Thermoelectric properties of ultrathin silicon nanowires.
	\newblock {\em Phys. Rev. B}, 86:115328, 2012.
	
	\bibitem{McGaugheyAPL12}
	Alan J.~H. McGaughey and Ankit Jain.
	\newblock Nanostructure thermal conductivity prediction by monte carlo sampling
	of phonon free paths.
	\newblock {\em Appl. Phys. Lett.}, 100(6):061911, 2012.
	
	\bibitem{LiJAP13}
	Wu~Li and Natalio Mingo.
	\newblock Alloy enhanced anisotropy in the thermal conductivity of sixge1âˆ’x
	nanowires.
	\newblock {\em J. Appl. Phys.}, 114(5):054307, 2013.
	
	\bibitem{KukitaJAP13}
	K.~Kukita and Y.~Kamakura.
	\newblock Monte carlo simulation of phonon transport in silicon including a
	realistic dispersion relation.
	\newblock {\em J. Appl. Phys.}, 114(15):154312, 2013.
	
	\bibitem{MeiJAP14}
	S.~Mei, L.~N. Maurer, Z.~Aksamija, and I.~Knezevic.
	\newblock Full-dispersion monte carlo simulation of phonon transport in
	micron-sized graphene nanoribbons.
	\newblock {\em J. Appl. Phys.}, 116(16):164307, 2014.
	
	\bibitem{WeberABC2}
	W.~Weber.
	\newblock Adiabatic bond charge model for the phonons in diamond, {Si, Ge, and
		$\alpha$-Sn}.
	\newblock {\em Phys. Rev. B}, 15:4789--4803, 1977.
	
	\bibitem{LarkinJAP13}
	Jason~M. Larkin and Alan J.~H. McGaughey.
	\newblock Predicting alloy vibrational mode properties using lattice dynamics
	calculations, molecular dynamics simulations, and the virtual crystal
	approximation.
	\newblock {\em J. Appl. Phys.}, 114(2):023507, 2013.
	
	\bibitem{UpadhyayaJMR15}
	Meenakshi Upadhyaya, Seyedeh~Nazanin Khatami, and Zlatan Aksamija.
	\newblock Engineering thermal transport in sige-based nanostructures for
	thermoelectric applications.
	\newblock {\em J. Mater. Res.}, FirstView:1--14, 2015.
	
	\bibitem{KhatamiPRApp16}
	S.~N. Khatami and Z.~Aksamija.
	\newblock Lattice thermal conductivity of the binary and ternary group-iv
	alloys si-sn, ge-sn, and si-ge-sn.
	\newblock {\em Phys. Rev. Applied}, 6:014015, 2016.
	
	\bibitem{MaurerAPL15}
	L.~N. Maurer, Z.~Aksamija, E.~B. Ramayya, A.~H. Davoody, and I.~Knezevic.
	\newblock Universal features of phonon transport in nanowires with correlated
	surface roughness.
	\newblock {\em Appl. Phys. Lett.}, 106(13):133108, 2015.
	
	\bibitem{GoodnickPRB85}
	S.~M. Goodnick, D.~K. Ferry, C.~W. Wilmsen, Z.~Liliental, D.~Fathy, and O.~L.
	Krivanek.
	\newblock Surface roughness at the si(100)-si02 interface.
	\newblock {\em Phys. Rev. B}, 32(12):8171, 1985.
	
	\bibitem{DharAdvPhys08}
	Abhishek Dhar.
	\newblock Heat transport in low-dimensional systems.
	\newblock {\em Advances in Physics}, 57(5):457--537, 2008.
	
	\bibitem{LepriPhysRep03}
	Stefano Lepri, Roberto Livi, and Antonio Politi.
	\newblock Thermal conduction in classical low-dimensional lattices.
	\newblock {\em Phys. Reports}, 377(1):1 -- 80, 2003.
	
	\bibitem{MaiPRL07}
	Trieu Mai, Abhishek Dhar, and Onuttom Narayan.
	\newblock Equilibration and universal heat conduction in fermi-pasta-ulam
	chains.
	\newblock {\em Phys. Rev. Lett.}, 98:184301, 2007.
	
	\bibitem{VermeerschAPL16}
	Bjorn Vermeersch, Jesús Carrete, and Natalio Mingo.
	\newblock Cross-plane heat conduction in thin films with ab-initio phonon
	dispersions and scattering rates.
	\newblock {\em Appl. Phys. Lett.}, 108(19):193104, 2016.
	
	\bibitem{YangNT10}
	Nuo Yang, Gang Zhang, and Baowen Li.
	\newblock Violation of fourier's law and anomalous heat diffusion in silicon
	nanowires.
	\newblock {\em Nano Today}, 5(2):85 -- 90, 2010.
	
	\bibitem{SaitoPRL10}
	Keiji Saito and Abhishek Dhar.
	\newblock Heat conduction in a three dimensional anharmonic crystal.
	\newblock {\em Phys. Rev. Lett.}, 104:040601, 2010.
	
	\bibitem{BaeNC13}
	Myung-Ho Bae, Zuanyi Li, Zlatan Aksamija, Pierre~N Martin, Feng Xiong,
	Zhun-Yong Ong, Irena Knezevic, and Eric Pop.
	\newblock Ballistic to diffusive crossover of heat flow in graphene ribbons.
	\newblock {\em Nat. Commun.}, 4:1734, 2013.
	
	\bibitem{ViningJAP91_2}
	Cronin~B. Vining.
	\newblock A model for the high-temperature transport properties of heavily
	doped n-type silicon-germanium alloys.
	\newblock {\em J. Appl. Phys.}, 69:331--341, 1991.
	
	\bibitem{DharPRE13}
	Abhishek Dhar, Keiji Saito, and Bernard Derrida.
	\newblock Exact solution of a l\'evy walk model for anomalous heat transport.
	\newblock {\em Phys. Rev. E}, 87:010103, 2013.
	
	\bibitem{ZaburdaevRMP15}
	V.~Zaburdaev, S.~Denisov, and J.~Klafter.
	\newblock L\'evy walks.
	\newblock {\em Rev. Mod. Phys.}, 87:483--530, 2015.
	
	\bibitem{CiprianiPRL05}
	P.~Cipriani, S.~Denisov, and A.~Politi.
	\newblock From anomalous energy diffusion to levy walks and heat conductivity
	in one-dimensional systems.
	\newblock {\em Phys. Rev. Lett.}, 94:244301, 2005.
	
	\bibitem{DenisovPRL03}
	S.~Denisov, J.~Klafter, and M.~Urbakh.
	\newblock Dynamical heat channels.
	\newblock {\em Phys. Rev. Lett.}, 91:194301, 2003.
	
	\bibitem{LiuPRL14}
	Sha Liu, Peter H\"anggi, Nianbei Li, Jie Ren, and Baowen Li.
	\newblock Anomalous heat diffusion.
	\newblock {\em Phys. Rev. Lett.}, 112:040601, 2014.
	
	\bibitem{ShlesingerNat93}
	Michael~F. Shlesinger, George~M. Zaslavsky, and Joseph Klafter.
	\newblock Strange kinetics.
	\newblock {\em Nature}, 363:31--37, 1993.
	
	\bibitem{LiNJP15}
	Yunyun Li, Sha Liu, Nianbei Li, Peter Hänggi, and Baowen Li.
	\newblock 1d momentum-conserving systems: the conundrum of anomalous versus
	normal heat transport.
	\newblock {\em New J. Phys.}, 17(4):043064, 2015.
	
	\bibitem{MorelliPRB02}
	D.~T. Morelli, J.~P. Heremans, and G.~A. Slack.
	\newblock Estimation of the isotope effect on the lattice thermal conductivity
	of group {IV} and group {III-V} semiconductors.
	\newblock {\em Phys. Rev. B}, 66(19):195304, 2002.
	
	\bibitem{KongPRB09}
	B.~D. Kong, S.~Paul, M.~Buongiorno Nardelli, and K.~W. Kim.
	\newblock First-principles analysis of lattice thermal conductivity in
	monolayer and bilayer graphene.
	\newblock {\em Phys. Rev. B}, 80(3):033406, 2009.
	
	\bibitem{KlemensCarbon94}
	P.G. Klemens and D.F. Pedraza.
	\newblock Thermal conductivity of graphite in the basal plane.
	\newblock {\em Carbon}, 32:735 -- 741, 1994.
	
	\bibitem{SlackSSPv34}
	G.~Slack.
	\newblock {\em Solid State Physics, vol. 34}.
	\newblock Academic Press, New York, NY, 1979.
	
	\bibitem{GargPRL11}
	Jivtesh Garg, Nicola Bonini, Boris Kozinsky, and Nicola Marzari.
	\newblock Role of disorder and anharmonicity in the thermal conductivity of
	silicon-germanium alloys: A first-principles study.
	\newblock {\em Phys. Rev. Lett.}, 106:045901, Jan 2011.
	
	\bibitem{WeiPRL93}
	Lanhua Wei, P.~K. Kuo, R.~L. Thomas, T.~R. Anthony, and W.~F. Banholzer.
	\newblock Thermal conductivity of isotopically modified single crystal diamond.
	\newblock {\em Phys. Rev. Lett.}, 70(24):3764--3767, 1993.
	
\end{thebibliography}


\end{document}